\documentclass[conference]{IEEEtran}
\usepackage{amsmath,amsfonts}
\usepackage{amssymb}
\usepackage{amsthm}
\usepackage[hidelinks]{hyperref}
\usepackage{algorithmic}
\usepackage{graphicx}
\usepackage{textcomp}
\usepackage{xcolor}
\usepackage{multirow}
\usepackage{booktabs}
\usepackage{tikz}
\usetikzlibrary{shapes,shapes.geometric,shapes.symbols,arrows,arrows.meta,positioning,fit,calc,backgrounds,decorations.pathreplacing}
\usepackage{xspace}
\usepackage{seqsplit}
\usepackage{wrapfig}

\newcommand{\NAME}{\textsf{Mirage}\xspace}
\newcommand{\GRa}{$11.6$}
\newcommand{\GRb}{$12.0$}
\newcommand{\GRc}{$12.1$}
\newcommand{\GRd}{$11.1$}
\newcommand{\GRn}{$5$}

\newcommand{\ITLdsNODEFttft}{$0.97$}
\newcommand{\ITLdsNODEFitl}{$0.81$}
\newcommand{\ITLdsNODEFboth}{$0.98$}
\newcommand{\ITLdsNODEFpt}{$0.96$}
\newcommand{\ITLdsMIRttft}{$0.60$}
\newcommand{\ITLdsMIRitl}{$0.63$}
\newcommand{\ITLdsMIRboth}{$0.60$}
\newcommand{\ITLdsMIRpt}{$0.56$}
\newcommand{\repo}[1]{\texttt{\seqsplit{#1}}}
\newtheorem{theorem}{Theorem}
\newtheorem{lemma}{Lemma}
\newtheorem{assumption}{Assumption}
\newtheorem{corollary}{Corollary}
\newtheorem{proposition}{Proposition}
\DeclareMathOperator{\TV}{TV}

\newcommand{\hit}{\mathsf{hit}}
\newcommand{\miss}{\mathsf{miss}}
\newcommand{\Lo}{\mathcal{L}}
\newcommand{\Tr}{\mathcal{T}}
\newcommand{\deq}{\delta_{\mathsf{eq}}}

\providecommand{\Description}[1]{}
\usepackage{microtype}

\begin{document}
\title{\NAME: Provable Defense Against LLM KV Cache Timing Side Channels with Topology-Aware Isolation}

\author{
\IEEEauthorblockN{Xiaobei Yan}
\IEEEauthorblockA{Nanyang Technological University\\
\href{mailto:xiaobei002@e.ntu.edu.sg}{xiaobei002@e.ntu.edu.sg}}
\and
\IEEEauthorblockN{Tianwei Zhang}
\IEEEauthorblockA{Nanyang Technological University\\
\href{mailto:tianwei.zhang@ntu.edu.sg}{tianwei.zhang@ntu.edu.sg}}
}

\hypersetup{pdfauthor={Xiaobei Yan and Tianwei Zhang}}
\maketitle
\begin{abstract}
Large Language Models (LLMs) are increasingly deployed as shared multi-tenant services, relying on advanced optimizations like KV cache and prefix sharing to improve throughput and reduce latency. However, these optimizations introduce a critical vulnerability: attackers can exploit time-to-first-token (TTFT) variations between KV cache hits and misses to infer sensitive prompt tokens via prefix-matching side-channel attacks. Existing defenses indiscriminately restrict cross-tenant cache sharing, which incurs prohibitive performance overhead and negates the fundamental efficiency benefits of the KV cache.

We propose \NAME, a topology-aware emission obfuscation framework that provably bounds timing leakage without sacrificing the performance gains of cache sharing. Instead of stalling the GPU or forcing redundant recomputations, \NAME delays token emission at the gateway so the attacker observes a natural cache miss distribution, while the GPU seamlessly continues serving other requests at full speed. \NAME uses a Privacy Odometer that persistently tracks the topological novelty of prefix fan-out expansions to distinguish aggressive probing from benign traffic. Once a prefix exceeds a safety threshold, \NAME engages a two-phase policy that uses a dynamic estimator to pad the TTFT to a calibrated cache-miss baseline. On a 500-sample public corpus across three models, \NAME suppresses the timing side-channel to near the random-guessing floor, while maintaining a significantly lower TTFT than defense baselines. We further show that \NAME withstands a strong adaptive adversary and the same principle carries over to semantic caching layers.
\end{abstract}
\IEEEpeerreviewmaketitle
\section{Introduction}

Large language models (LLMs) are increasingly deployed as shared cloud infrastructure, driving their adoption in privacy-sensitive fields like medical~\cite{llm_medical_survey}, finance~\cite{wu2023bloomberggpt}, and coding assistants~\cite{copilot}. This paradigm shift makes inference latency a first-order system constraint. Users expect interactive responsiveness, while providers must efficiently amortize scarce GPU compute and memory across concurrent requests. While early optimizations focused on weight quantization~\cite{yao2022zeroquant,wei2022outlier,xiao2023smoothquant}, model compression~\cite{wang2024model}, and backend enhancements~\cite{dao2022flashattention}, modern serving engines increasingly rely on key-value (KV) cache management and prefix reuse to eliminate redundant computations.
By organizing cached prefixes through block hashes or radix-tree structures, systems like vLLM and SGLang allow subsequent requests sharing initial tokens to bypass the expensive prefill phase and directly reuse cached KV tensors~\cite{kwon2023pagedattention,zheng2024sglang}. Today, prefix sharing serves as a foundational performance primitive across modern APIs, significantly accelerating workloads that rely on long prompts, repeated templates, retrieval-augmented generation (RAG), and multi-turn conversations~\cite{hydragen,chunk-attention,openai2024promptcaching,anthropic2024promptcaching,deepseek2024contextcaching}.

Unfortunately, the KV cache sharing scheme exposes a timing oracle. A request whose prefix hits the shared KV cache reaches the first generated token much faster than a request whose prefix must be computed from scratch. Prior work has shown that an attacker can exploit this time-to-first-token (TTFT) gap using only ordinary API access: the attacker sends many guessing probes, measures whether the probe behaves like a cache hit, and repeats the process to recover private prompts, system instructions, or fields from another user's input~\cite{early-bird,input-snatch,prompt-peek,audit-PSA}. This risk is especially acute in multi-tenant deployments because sensitive content such as medical questions, legal facts, enterprise documents, and proprietary agent prompts may coexist with performance-oriented cross-user cache sharing.

Existing mitigations leave an uncomfortable trade-off. Per-user cache partitioning and cache salts prevent cross-tenant cache hits, but also remove the cross-tenant reuse that prefix caching was designed to provide~\cite{cache-partitioning,vllmCacheSalt}. Selective-sharing systems such as SafeKV~\cite{safeKV} reduce this cost by classifying which cache entries are sensitive, but they still respond by \emph{isolating} all entries classified as sensitive. Therefore, a genuine user of a sensitive prefix loses all cache reuse, the very benefit caching provides. Besides, their security relies entirely on an imperfect classifier: a false negative leaks the sensitive prefix completely, while a false positive needlessly forfeits cache reuse for benign traffic. Fundamentally, these defenses label content while ignoring attackers' \emph{behaviors}, so they cannot distinguish a benign user reading a sensitive document from an adversary hammering it with candidate probes. 

A seemingly simpler alternative is timing obfuscation: add delay or jitter to suspicious first-token emissions. However, this introduces several unique challenges: First, indiscriminately adding noise negates the latency-reduction benefits of the KV cache. As pointed out in prior studies~\cite{safeKV}, completely masking the timing channel requires padding all cache-hit response time to the upper tail of the cache-miss distribution. This directly obliterates the substantial latency reductions (often up to 73\%) typically provided by prefix caching, rendering the underlying performance optimization practically meaningless. Second, because natural cache miss latencies fluctuate dynamically with server load, sampling an artificial delay that convincingly mimics the real distribution is non-trivial. Third, obfuscation must be applied with care in streaming APIs: if only the first token is held back while the engine keeps decoding, the buffered tokens are released in a burst, so a correct implementation shifts the entire emission stream rather than its first token alone.
Furthermore, establishing rigorous mathematical bounds on an attacker's residual advantage when facing such obfuscation remains an open and necessary challenge. Finally, designing this defense to integrate easily into production-grade serving environments as a plug-and-play module with minimal revision to current frameworks is a critical practical consideration.

To address these challenges, this paper presents \NAME, a theoretically bounded defense that formally limits residual timing leakage while preserving the compute savings of prefix caching. \NAME is based on a simple observation: cache probing attacks are not arbitrary traffic: the attacker must target a sensitive prefix. Because the unknown private text possesses high information entropy, while each timing probe reveals at most a few bits (the hit/miss verdict), the attacker is mathematically forced to issue a massive number of distinct candidate guessing probes beneath that exact same prefix. In the prefix cache's logical topology, this creates a distinctive high-fan-out pattern: many novel outgoing branches stemming from a single metered node. \NAME turns this fundamental attack requirement into a control-plane budget. It tracks how many topologically novel continuations a sensitive prefix has accumulated, decays this state over time, and moves the prefix into an emission lockdown phase once the safety budget is exhausted. Because this signal sits on the critical path of every request, it is kept deliberately lightweight, and \NAME shows that a single, cheap behavioral signal is sufficient to close the channel.
When a request enters lockdown, \NAME does not stop the GPU. Instead, it performs \textit{time shifting} at the API emission layer. After the first token is produced, \NAME samples a target latency from a calibrated cache-miss distribution and computes a constant offset between the true token arrival time and the target emission time. The gateway then applies this same offset to every chunk in the stream. As a result, the attacker observes a natural miss-like TTFT and natural inter-token latency (ITL). The GPU remains free to continue serving other work at full speed, while the client-visible trace is shifted as a whole.

We implement \NAME in SGLang and evaluate it against three representative KV-cache timing attacks across Qwen, DeepSeek, and Llama-family models. \NAME reduces the success rate of all three attack baselines to near the random-guessing floor while preserving benign public-prefix latency. By contrast, isolation-style defenses forfeit shared-prefix reuse and inflate benign median TTFT by roughly $45\%$ on the 14B model at moderate load, and several-fold on smaller models as load rises.

In summary, this paper makes the following contributions:
\begin{itemize}
    \item \textbf{Metering fan-out rather than classifying secrets.} Any timing attacker recovering a high-entropy secret must open many novel branches beneath the same prefix; we prove a Fano-style lower bound on this fan-out and make it the primary detection signal. A coarse content-sensitivity weight only scales how fast a prefix accrues risk, so \NAME acts on the attacker's unavoidable exploration behavior instead of deciding per request whether a prompt contains a secret.
    \item \textbf{Isolation-level security without sacrificing cache reuse.} \NAME serves all prefixes from the shared KV cache at full GPU speed, applying stream-level time-shifting at the gateway only during lockdown. By eliminating the redundant recomputations forced by isolation-based solutions, \NAME retains up to $1.6\times$ higher serving throughput, maintains near-zero memory overhead, and confines latency penalties strictly to sensitive traffic that is under active probing, all with provable per-request latency and capacity bounds (Appendix~\ref{sec:perf-bounds}).
    \item \textbf{Stream-level timing equalization with provable bounds.} We theoretically bound the attacker's residual advantage by an equalization gap $\deq$ and design a dynamic tail-quantile clamp that shifts the entire token stream. We empirically show this eliminates whole-trace timing leaks, keeping state-of-the-art multivariate and ITL-aware distinguishers at chance across three model scales and batched loads.
    \item \textbf{A plug-and-play realization.} Implemented as a module over unmodified SGLang, one \NAME policy drives three published timing attacks to the guessing floor on three model families, resists a white-box adaptive adversary, and carries over to semantic caches.
\end{itemize}

\section{Background}

\subsection{LLM Inference and Streaming Latency}
Transformer-based LLMs generate text autoregressively using self-attention over query, key, and value vectors~\cite{vaswani2017attention}. Serving a request has two stages. In the \textit{prefill} stage, the model processes the entire input prompt in parallel, computes the per-layer key/value tensors for all prompt tokens, and produces the first output token. In the \textit{decode} stage, the model generates subsequent tokens one at a time; each step computes the key/value tensors for the newest token and attends over the previous context. Modern serving systems treat these stages differently because prefill is compute-heavy and prompt-length dependent, while decode is latency-sensitive and sequential~\cite{zhong2024distserve}.

Interactive LLM APIs commonly stream tokens as they are generated. Two latency metrics are therefore visible to clients. TTFT is the elapsed time from request submission to the first emitted token, and mostly captures queueing plus prefill work. \textit{Inter-token latency} (ITL), also called time per output token, is the gap between consecutive emitted tokens and mostly captures decode cadence. For a completion of $N$ generated tokens, user-visible latency can be decomposed as $\mathrm{TTFT} + \sum_{i=2}^{N}\mathrm{ITL}_i$. 

\subsection{KV Cache and Prefix Reuse}
The KV cache stores the key and value tensors computed for previous tokens so that later attention steps can reuse them instead of recomputing the entire context. This reduces the per-step decoding cost from repeatedly processing all prior tokens to processing only the newly generated token, which is essential for long contexts and multi-turn conversations~\cite{pope2023efficiently}. However, KV tensors are large and must be managed carefully across batches, active requests, and memory tiers. PagedAttention in vLLMs, for example, partitions KV memory into blocks and uses metadata to support flexible allocation and reuse~\cite{kwon2023pagedattention}.

Prefix caching extends KV reuse across requests. Because causal attention makes a token's KV tensors depend on all preceding tokens, two requests can share cached tensors only for an identical prefix starting at the first token. If one request begins with the same token sequence as an earlier request, the serving engine can reuse the cached prefix and prefill only the unmatched suffix. This optimization is widely adopted in open-source systems and commercial APIs because many workloads repeatedly use system prompts, few-shot examples, RAG templates, long documents, or conversation history~\cite{zheng2024sglang,hydragen,chunk-attention,openai2024promptcaching,anthropic2024promptcaching,deepseek2024contextcaching}.

\subsection{Timing-based Prefix-Probing Attacks}
To facilitate prefix sharing, serving systems employ dedicated index structures to map input tokens to their corresponding KV tensors. For example, SGLang utilizes a Radix Tree (RadixAttention) to manage cached prefixes~\cite{zheng2024sglang}, while vLLM relies on block-level hashing mechanisms~\cite{kwon2023pagedattention}. Though structurally different, both designs yield the same operational outcome: an incoming request either matches a cached prefix to bypass redundant computations, or misses and triggers a full prompt recomputation. Crucially, these cache entries reside in the volatile data plane (e.g., GPU memory) and are managed by heuristic eviction policies (like LRU) optimizing for performance rather than privacy.

This architectural choice inadvertently exposes a highly observable timing oracle~\cite{prompt-peek, early-bird, input-snatch}. Because a prefix cache hit removes intensive prefill computation from the critical path, the TTFT drops significantly compared to a cache miss. In a multi-tenant environment, an attacker can exploit this discrepancy to reverse-engineer co-tenants' private data. Specifically, for a targeted prefix $P$, the attacker issues probes of the form $P \Vert c_i \Vert R$, where $c_i$ is a candidate token or phrase and $R$ is an optional suffix that keeps the request syntactically valid. If $P \Vert c_i$ matches a victim-cached continuation, the backend can reuse a longer cached prefix and the attacker's TTFT is drawn from a faster hit-like distribution; otherwise, the backend recomputes more of the prompt and the TTFT is drawn from a slower miss-like distribution. Repeating this test over many candidates~\cite{prompt-peek,input-snatch,early-bird} recovers the secret one token or span at a time.

\section{Threat Model and Motivation}
\label{sec:threat-model}

\subsection{Threat Model}
We consider a multi-tenant LLM serving backend that enables cross-request prefix caching for performance: the backend may reuse KV tensors when an incoming prompt shares a token prefix with a previous request. This is the setting studied by recent KV-cache leakage attacks and defenses~\cite{prompt-peek,input-snatch,early-bird,cache-partitioning,safeKV}. We assume a well-functioning and secure operating system or hypervisor, and that standard transport security and authentication are in place; the leakage considered here arises solely from API-visible timing, not from plaintext network exposure.

\noindent\textbf{Secrets and Attacker Objective.}
The victim's secret is a token sequence that should not be inferable by another tenant, such as a system prompt or a span of a private document. We focus on the incremental reconstruction problem used by prior attacks: given a known or guessed prefix $P$, the attacker tries to identify the next secret token or short span that follows $P$. The attacker may know public prompt templates, tokenizer behavior, or domain context, but not the victim's exact private continuation. A successful attack can leverage timing observations to distinguish correct continuation from candidate alternatives with probability significantly above guessing.

\noindent\textbf{Adversary Capabilities.}
The attacker is an ordinary black-box API client. He can create one or more accounts, issue chosen prompts, replay prompts, send probes concurrently or in carefully timed batches, and adapt later probes based on earlier timing measurements. He can observe the full streaming trace returned to his own requests: request submission time, TTFT, and the inter-token gaps of subsequent chunks. We grant the whole trace rather than TTFT alone so the evaluation also accounts for any residual signal in the post-TTFT emissions.

We also allow the attacker to directly flush the cache or create cache pressure by sending unrelated dummy requests, which triggers ordinary LRU-style eviction in the GPU-resident KV cache. Furthermore, the attacker possesses domain knowledge and can fine-tune his local auxiliary models to optimize candidate generation and maximize attack efficiency. However, the attacker has no privileged server access: he cannot read other users' prompts, inspect KV tensors, tamper with the control plane, etc.

\noindent\textbf{Defense Goals.}
Our goal is to bound the number of informative novel probes available for a sensitive prefix and, once that budget is exhausted, make the client-visible emission trace independent of whether the probe hits a victim's cached prefix. The protected observation is the whole streaming trace, not only TTFT: after lockdown, both the first-token emission time and subsequent inter-token latencies should be consistent with a natural miss-like execution.

Our solution does not claim to hide all information in the first grace-period probes; instead, it sets the topology budget below the number of novel candidates needed for reliable extraction. It also does not address attacks that require privileged access, direct KV cache content or model-output leakage, semantic leakage unrelated to prefix-cache timing, token-count leakage, or hardware and microarchitectural side channels outside the API timing surface. Public templates and explicitly trusted sharing groups are considered safe cache-sharing domains and may bypass our monitoring policy.

\subsection{Motivation}
\label{sec:motivation}
Three observations about the timing side channel motivate our design.
First, the channel is narrow but the secret is wide, and this asymmetry makes the attacker's \emph{behavior} distinguishable. A single probe reveals at most a few bits (the hit/miss verdict plus a coarse magnitude signal), while the secrets worth stealing carry far more entropy than that. Whatever candidate strategy the attacker adopts, it must therefore accumulate its answer across many distinct continuations under the same prefix, a sustained fan-out that no benign user of that prefix reproduces. The attack is thus visible in the shape of the traffic itself, before any judgment about which content is secret, and a defense can key on this signal instead of on content labels.

Second, the attacker observes latency, not computation under our threat model, which admits a cheaper equivalent of isolation. Existing defenses close the channel by actually withholding reuse: a protected prefix is recomputed from scratch for each trust boundary, which works against the very purpose of the KV cache. But since the client sees only timing, the same view can be produced without the recomputation: keeping serving the prefix from the shared cache in the backend, and delaying the emission so the response looks like a miss in the frontend. If the added delay follows the same distribution as a genuine miss, the two constructions are indistinguishable from the attacker's viewpoint, yet the GPU time that recomputation would burn is freed for other pending requests. In the worst case, where every request of a prefix ends up padded, this degrades to the same latency as isolation; in every other case, the reclaimed compute is pure gain.

Third, a defense is only deployable if it respects the serving stack it protects. Inference engines evolve quickly, so the defense should not modify the engine's scheduler, cache structures, or kernels; and since its logic runs on the critical path of every request in a latency-sensitive service, that logic must itself be near-free. This pushes the design toward a self-contained control-plane layer: light bookkeeping and emission control at the gateway, no GPU work, and no modification to the underlying serving engine.

\section{Methodology}
\label{sec:method}

\subsection{Design Overview}
\NAME is a stream-level defense for shared-prefix serving. It does not prevent prefix caching or stall the GPU when a suspicious request appears; instead, it separates privacy state tracking from physical execution and controls only the emitted timing. Three components work together: a \emph{behavioral} detector (the Privacy Odometer) that meters the attacker's unavoidable fan-out instead of guessing which content is secret; a \emph{decoupled} emitter (the gateway time-shift) that hides the hit/miss bit by shifting the whole client-visible stream while the GPU keeps running; and a \emph{calibrated} target that drives a padded hit toward statistical indistinguishability from a genuine miss, up to a small residual we measure ($\deq$). 

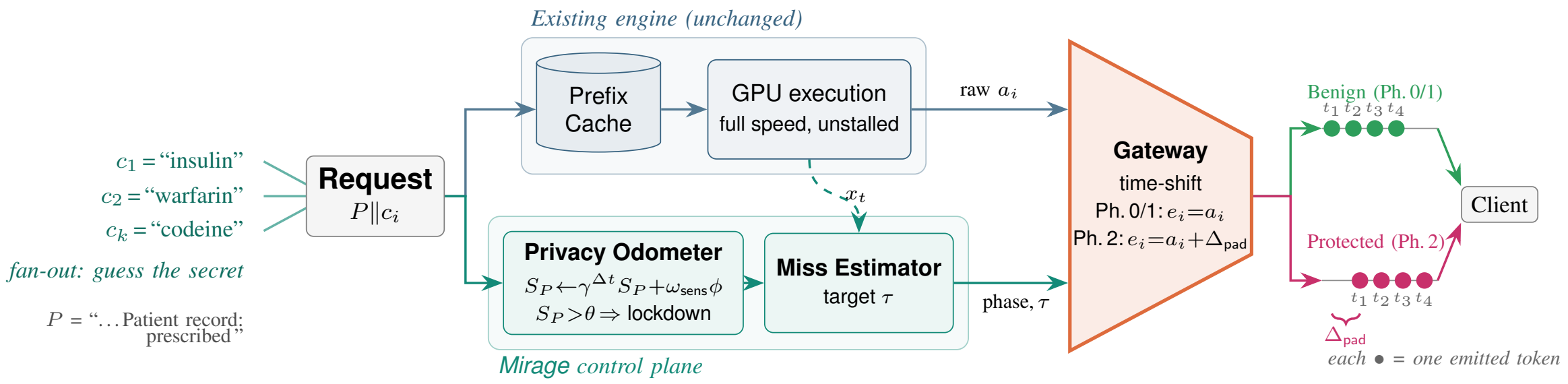
\begin{figure*}[t]
\centering
\definecolor{engblue}{HTML}{35617F}
\definecolor{mirteal}{HTML}{138A78}
\definecolor{gwcoral}{HTML}{DE6B3C}
\definecolor{okgreen}{HTML}{2E9E5B}
\definecolor{protrose}{HTML}{C8306B}
\resizebox{0.85\linewidth}{!}{\begin{tikzpicture}[
  font=\sffamily\normalsize,
  >={Stealth[length=2.6mm]},
  eng/.style={draw=engblue, fill=engblue!8, rounded corners=3pt, minimum height=12mm, inner sep=4pt, align=center},
  mir/.style={draw=mirteal, fill=mirteal!8, rounded corners=3pt, minimum height=12mm, inner sep=4pt, align=center},
  flow/.style={->, thick, draw=engblue!85},
  cflow/.style={->, thick, draw=mirteal},
  dot/.style={circle, fill=okgreen, inner sep=2pt},
  pdot/.style={circle, fill=protrose, inner sep=2pt},
]
\foreach \y/\lab/\drug in {0.42/{$c_1$}/{``insulin''},0.0/{$c_2$}/{``warfarin''},-0.42/{$c_k$}/{``codeine''}}{
  \node[font=\footnotesize\color{mirteal!80!black}, anchor=east] at (-1.62,\y) {\lab\,=\,\drug};
  \draw[mirteal!65, thick] (-1.5,\y) -- (-0.72,0);}
\node[font=\footnotesize\itshape\color{mirteal!80!black}, anchor=east] at (-1.62,-0.92) {fan-out: guess the secret};
\node[draw=black!50, rounded corners=2.5pt, fill=black!4, align=center, inner sep=4pt] (req) at (-0.15,0) {\textbf{Request}\\[-1.5pt]\footnotesize $P\Vert c_i$};
\node[font=\scriptsize\color{black!70}, align=right, anchor=north east] at (-1.62,-1.28)
  {$P$ = ``\ldots Patient record:\\[-2.5pt]prescribed\,''};

\node[cylinder, shape border rotate=90, aspect=0.28, draw=engblue, fill=engblue!12,
      minimum width=15mm, minimum height=12mm, inner sep=1pt, align=center] (cache) at (2.55,1.05) {\footnotesize Prefix\\[-2.5pt]\footnotesize Cache};
\node[eng, minimum width=19mm] (gpu) at (5.1,1.05) {\footnotesize GPU execution\\[-2pt]\scriptsize full speed, unstalled};
\begin{scope}[on background layer]
  \node[draw=engblue!40, fill=engblue!4, rounded corners=4pt, fit=(cache)(gpu), inner sep=5pt] (engbox){};
\end{scope}
\node[anchor=west, font=\footnotesize\itshape\color{engblue}] at ([yshift=6pt]engbox.north west) {Existing engine (unchanged)};

\node[mir, minimum width=29mm] (odo) at (2.85,-1.05) {\footnotesize\textbf{Privacy Odometer}\\[-1pt]\scriptsize $S_P{\leftarrow}\gamma^{\Delta t}S_P{+}\omega_{\mathsf{sens}}\phi$\\[-2.5pt]\scriptsize $S_P{>}\theta\!\Rightarrow$ lockdown};
\node[mir, minimum width=19mm] (est) at (5.7,-1.05) {\footnotesize\textbf{Miss Estimator}\\[-2.5pt]\scriptsize target $\tau$};
\begin{scope}[on background layer]
  \node[draw=mirteal!40, fill=mirteal!4, rounded corners=4pt, fit=(odo)(est), inner sep=5pt] (mirbox){};
\end{scope}
\node[anchor=west, font=\footnotesize\itshape\color{mirteal}] at ([yshift=-6pt]mirbox.south west) {\NAME control plane};

\node[trapezium, trapezium angle=62, shape border rotate=270, draw=gwcoral, very thick,
      fill=gwcoral!13, minimum width=20mm, minimum height=19mm, align=center, inner sep=1pt]
      (gw) at (9.35,0) {\footnotesize\textbf{Gateway}\\[-1pt]\scriptsize time-shift\\[-2.5pt]\scriptsize Ph.\,0/1:\,$e_i{=}a_i$\\[-2.5pt]\scriptsize Ph.\,2:\,$e_i{=}a_i{+}\Delta_{\mathsf{pad}}$};

\draw[flow] (req.east) -- ++(0.25,0) |- (cache.west);
\draw[cflow] (req.east) -- ++(0.25,0) |- (odo.west);
\draw[flow] (cache) -- (gpu);
\draw[cflow] (odo) -- (est);
\draw[cflow, dashed] (gpu.south) .. controls +(0,-0.55) and +(0,0.55) .. node[right, font=\scriptsize, pos=0.5]{$x_t$} (est.north);
\draw[flow] (gpu.east) -- node[above, font=\scriptsize]{raw $a_i$} (gpu.east -| gw.west);
\draw[cflow] (est.east) -- node[below, font=\scriptsize, pos=0.55]{phase,\,$\tau$} (est.east -| gw.west);

\node[font=\scriptsize\color{okgreen}, anchor=west] at (11.0,1.24) {Benign (Ph.\,0/1)};
\draw[black!45] (11.3,0.82) -- (12.7,0.82);
\foreach \x/\t in {11.42/1,11.68/2,11.94/3,12.2/4} \node[dot, label={[font=\tiny,color=black!60,inner sep=1pt]above:$t_{\t}$}] at (\x,0.82){};
\node[font=\scriptsize\color{protrose}, anchor=west] at (11.0,-0.56) {Protected (Ph.\,2)};
\draw[black!45] (11.3,-1.02) -- (12.7,-1.02);
\foreach \x/\t in {11.76/1,12.02/2,12.28/3,12.54/4} \node[pdot, label={[font=\tiny,color=black!60,inner sep=1pt]below:$t_{\t}$}] at (\x,-1.02){};
\draw[decorate, decoration={brace, amplitude=3.5pt}, protrose, thick] (11.76,-1.42) -- (11.42,-1.42)
  node[midway, below, font=\scriptsize, yshift=-1pt]{$\Delta_{\mathsf{pad}}$};
\node[font=\scriptsize\itshape\color{black!60}, anchor=west] at (11.25,-1.95) {each $\bullet$ = one emitted token};
\draw[flow, okgreen] (gw.east) -- ++(0.45,0) |- (11.3,0.82);
\draw[flow, protrose] (gw.east) -- ++(0.45,0) |- (11.3,-1.02);
\node[draw=black!50, rounded corners=2pt, fill=black!4, font=\footnotesize] (client) at (13.45,-0.1) {Client};
\draw[flow, okgreen] (12.7,0.82) -- (client.north west);
\draw[flow, protrose] (12.7,-1.02) -- (client.south west);
\end{tikzpicture}}
\caption{An attacker appends candidate guesses $c_i$ to a sensitive prefix $P$ and times the reply: a cache hit reveals the match. \NAME's control plane (teal) meters this fan-out while the unchanged engine (blue) serves the shared cache at full speed; the gateway (coral) passes benign streams through and shifts every token of a locked-down stream by one offset $\Delta_{\mathsf{pad}}$ (rose), so timing no longer reveals the hit.}
\label{fig:arch}
\end{figure*}

As illustrated in Figure~\ref{fig:arch}, the architecture is a decoupled defense layer appended to an existing LLM engine and joined at a gateway: the backend GPU continues to process tokens at full speed, while the gateway independently shapes the client-visible timing trace. The workflow spans two lanes joined at a gateway. First, incoming client requests enter the unmodified \textit{Existing engine} (the data plane), comprising the \textit{Prefix Cache} and \textit{GPU execution} core. The request is evaluated against the native radix tree to maximize cache hits and skip redundant prefill computations. In parallel, the request is evaluated by \NAME's decoupled \textit{control plane}. The \textit{Privacy Odometer} meters the topological novelty of the request's continuation beneath the sensitive prefix. It updates a time-decayed risk score $S_P$ and issues a Phase Flag (Phase 0, 1, or 2) dictating the defense posture. Operating alongside it, the \textit{Miss Estimator} maintains a dual-track exponential moving average (EMA) using actual TTFT feedback ($x_t$) from the GPU, continuously sampling a dynamic, load-aware cache-miss target latency ($\tau$). Finally, the \textit{Gateway} intercepts the raw tokens and acts as the decisive traffic router. Based on the Phase Flag from the odometer, it splits emissions into two diverging paths:
\begin{itemize}
    \item \textbf{Benign/Grace Stream (Phase 0/1):} Public traffic (i.e., general-purpose queries, standard instruction templates, or any request assigned a zero sensitivity weight) alongside the initial grace-period queries of sensitive prefixes, simply pass through ($e_i = a_i$) with unaltered fast TTFTs, retaining 100\% of prefix cache's performance benefits.
    \item \textbf{Protected Stream (Phase 2):} Once the probing budget is exhausted, the gateway triggers an emission lockdown. It computes a constant offset ($\Delta_{\mathsf{pad}}$) bridging the actual execution time and the dynamic miss target ($\tau$). The gateway then shifts every streamed chunk by this identical offset ($e_i = a_i + \Delta_{\mathsf{pad}}$). As a result, the attacker observes a natural miss-like TTFT while the subsequent inter-token latencies remain uncompressed and natural.
\end{itemize}

The architectural separation between the defense layer and native engine is specifically designed for robust security. The underlying GPU may evict KV blocks under memory pressure, but the \NAME odometer remains securely in CPU memory or a distributed store, decaying only according to its own logical policy. This completely decouples the privacy state from the physical cache, preventing an attacker from using cache-flush floods to reset the defense and launch subsequent probes.

\subsection{Privacy Odometer}

The Privacy Odometer is a control-plane counter that tracks the \emph{logical topology} of prefix reuse. Instead of metering physical cache hits (which would allow attackers to test incorrect candidates for free until they succeed), the odometer increments a risk score whenever a request creates a previously unseen continuation (a ``topologically novel branch'') beneath a metered prefix. Exact retries open no new branches, so they are never penalized.

For a request $r_j$ arriving at time $t_j$, \NAME decomposes the prompt into a metered prefix $P$ and a continuation branch $b_j$. The prefix $P$ acts as a fixed metering anchor that does not move as the prompt grows. The branch $b_j$ is a cryptographic hash of the \emph{entire} continuation that follows $P$. This full-continuation hashing ensures that any attempt to explore different candidates is robustly registered as a novel branch, even if injected into the middle of a template ($P \Vert \mathsf{candidate} \Vert \mathsf{suffix}$). Genuine retries naturally hash to the same branch.

Let $B_P$ be the persistent set of branches already observed below $P$. The topological novelty term is defined as:
\begin{equation}
\phi_{\mathsf{topo}}(r_j)=
\begin{cases}
1, & b_j \notin B_P,\\
0, & b_j \in B_P.
\end{cases}
\end{equation}

\noindent\textbf{Trust Scope as an Accounting Key.}
Before calculating the risk score, \NAME assigns the request to a \emph{Trust Scope ($c$)}. A scope acts strictly as a logical \emph{accounting key} (a behavioral risk ledger) rather than a physical cache boundary. In practice, the system maintains a separate risk ledger for each uniquely authenticated user. For example, if Dr.\,A and Dr.\,B both query the same sensitive patient record, \NAME serves both requests using the exact same physical KV cache in the GPU to maximize performance. However, their distinct probing behaviors are metered in completely separate scope ledgers. Consequently, the natural diversity of queries from many benign doctors legitimately sharing one record can never pool together into a single false alarm.

To defend against Sybil attacks, unauthenticated or anonymous traffic is collapsed into a single pooled scope. An adversary cannot bypass the mechanism by generating thousands of fake connections or rotating IP addresses to acquire a fresh grace budget for each probe, as all their anonymous attempts will accumulate against the identical pooled ledger and swiftly trigger the defense.

\noindent\textbf{Risk Score Update.} 
After computing $\phi_{\mathsf{topo}}$, \NAME inserts $b_j$ into $B_P$ and updates the risk score for the specific $(P, c)$ pair:
\begin{equation}
\begin{aligned}
S_{P,c}(t_j)=\min\Big\{&S_{\max},\ \gamma^{t_j-t_{j-1}} S_{P,c}(t_{j-1})\\
&+\omega_{\mathsf{sens}}(P)\big(\phi_{\mathsf{topo}}(r_j)+\rho(1-\phi_{\mathsf{topo}}(r_j))\big)\Big\}.
\end{aligned}
\label{eq:score}
\end{equation}
Here, $\gamma\in(0,1]$ is the temporal decay factor, $\omega_{\mathsf{sens}}(P)$ is the prefix sensitivity weight, and $\rho\in[0,1]$ is an optional penalty for exact retries (we default to $\rho=0$). 

The score is capped at $S_{\max}$. This cap prevents Denial-of-Service (DoS) attacks where a malicious user floods a shared prefix to lock it permanently. According to Equation~\ref{eq:decay-time} in Section~\ref{sec:security}, a locked prefix is mathematically guaranteed to decay below the threshold within a bounded idle time. Even if an attacker intentionally floods a pooled scope to sustain a lock, the result is merely a temporary latency degradation. This degraded latency exactly matches the baseline penalty that strict isolation defenses (e.g., SafeKV, per-user partitioning, vLLM salt-scoped caching) impose unconditionally on \emph{every} cross-boundary access. In other words, such flooding degrades system availability for that prefix, but never breaches confidentiality.

Finally, this privacy state (the decayed score and compact branch set $B_P$) is deliberately decoupled from the GPU data plane. As the odometer lives securely in the control plane memory, attackers cannot reset their risk scores by intentionally triggering GPU LRU cache evictions or flushes.

\subsection{Sensitivity Gate and Gateway Multiplexer}
The gateway first evaluates a sensitivity weight $\omega_{\mathsf{sens}}(P)\in[0,1]$ for the metered prefix. In \NAME, a ``sensitive token'' is not meant to be an exhaustively enumerated secret vocabulary. It is any token, structured span, application route, or trust-boundary label indicating that continuations below the current prefix may encode private information. The weight is therefore a policy interface: $\omega_{\mathsf{sens}}(P)=0$ means public (i.e., general-purpose queries, standard instruction templates, or any request assigned a zero sensitivity weight) or trusted-sharing traffic, while $\omega_{\mathsf{sens}}(P)>0$ means the prefix is privacy-bearing and should be metered.

Our prototype instantiates this interface with a reproducible lexical and structural gate for the medical-consultation benchmark. It assigns high weight to protected domain contexts (\texttt{patient}, \texttt{clinical}, \texttt{healthcare}), names and surnames from public Census-style lists, medical conditions and medications, legal/financial terms, and structured identifiers such as email addresses, phone numbers, long string/numeric identifiers, or private-location contexts. Crucially, the security of \NAME does \emph{not} rest on this list being complete. The safe deployment posture is \emph{fail-closed}: a provider sets $\omega_{\mathsf{sens}}=1$ by default for an entire endpoint, tenant namespace, private document class, or system-prompt template, so every request on a private surface is metered without content inspection, and the lexical gate serves only as a whitelist that lowers $\omega_{\mathsf{sens}}$ for known-public templates to reclaim their cache speed. A secret that ``slips through'' the list is therefore still protected by the fan-out budget of its metered endpoint; content scanning is a performance optimization, never the foundation of the guarantee. Nor does fail-closed \emph{metering} mean fail-closed \emph{padding}: a metered request is watched, not slowed. Padding engages only once a scope's fan-out budget is exhausted, so benign traffic keeps the fast cache path while the behavioral signal decides when, and for whom, emission is equalized.

\NAME then uses a threshold $\theta$ to switch from a grace period to lockdown:
\begin{equation}
\mathsf{phase}(P,t)=
\begin{cases}
\mathsf{Phase\ 0}, & \omega_{\mathsf{sens}}(P)=0,\\
\mathsf{Phase\ 1}, & \omega_{\mathsf{sens}}(P)>0 \text{ and } S_P(t)\le \theta,\\
\mathsf{Phase\ 2}, & S_P(t)>\theta.
\end{cases}
\end{equation}
Phase 0 is a performance whitelist path: requests identified as non-sensitive by the application layer or sensitivity gate (e.g., standard public instruction templates, system prompts, or general-knowledge queries) bypass the odometer and receive no emission padding, so they retain the native fast-cache TTFT. Phase 1 is the grace period for metered prefixes; the request also passes through unpadded because the prefix has not yet consumed its topology budget. Neither phase adds any equalization delay; the only residual cost is a small, constant control-plane term (the sensitivity classification and gateway interception applied to every request, quantified in Section~\ref{sec:performance}), not a per-request padding cost. Phase 2 is lockdown: emissions for that prefix are shaped to a miss-like distribution to shut down the timing oracle channel. Since every novel branch increases the score by at least $\omega_{\mathsf{sens}}(P)$ when $\rho=0$, a rapid attacker can create at most

\begin{equation}
K_{\max}(P)=\left\lfloor \frac{\theta}{\omega_{\mathsf{sens}}(P)} \right\rfloor
\label{eq:kmax}
\end{equation}
distinct Phase-1 candidate branches before the prefix enters lockdown and every later probe is padded.

\noindent\textbf{Configurable Grace Budget.}
Let $G$ be the number of unpadded novel branches a sensitive prefix is allowed before lockdown. The budget $\theta$ takes the form $G\,\omega_{\mathsf{sens}}\le\theta<(G+1)\omega_{\mathsf{sens}}$, as each novel branch adds exactly $\omega_{\mathsf{sens}}$ to the score (Equation~\ref{eq:score}) with $\rho=0$, so after the $G$-th distinct candidate the score is $G\,\omega_{\mathsf{sens}}$. The left inequality ($\theta\ge G\,\omega_{\mathsf{sens}}$) keeps those first $G$ probes in Phase~1, and the right inequality ($\theta<(G+1)\omega_{\mathsf{sens}}$) forces the $(G{+}1)$-th probe, which lifts the score to $(G{+}1)\omega_{\mathsf{sens}}$, over the threshold and into lockdown. The two bounds together pin lockdown to exactly the $(G{+}1)$-th novel candidate. Our main configuration adopts $G=1$ by setting $\theta=1.0$ with $\omega_{\mathsf{sens}}=1.0$: the \emph{first} novel branch beneath a sensitive prefix evaluates to $S_P=1.0\le\theta$ and passes through Phase~1 with zero overhead, so the first genuine visitor of a sensitive template is never penalized; a rapid second candidate jumps the score to $S_P=2.0>\theta$ and immediately triggers lockdown. This is a deliberate trade-off. A single unpadded probe cannot resolve a high-entropy secret (by Equation~\ref{eq:union} below, its success is bounded by the prior guessing probability $p_B$), so \NAME trades an exploitable, unbounded timing channel for one bounded guess while preserving the benign first-access latency that isolation defenses destroy. A deployment that wants strictly zero first-visit leakage can instead set $G=0$ ($\theta<\omega_{\mathsf{sens}}$), padding the first sensitive access; Equation~\ref{eq:kmax} keeps either choice explicit and auditable. Section~\ref{sec:adaptive} evaluates the low-and-slow adversary that the grace budget admits.

\subsection{Emission Obfuscator}
Naive first-token padding is unsafe for streaming APIs. Suppose the first token is held for $\Delta$ seconds while the backend continues decoding. Several later chunks may accumulate behind the sleep and then be released in a burst, producing abnormally small ITLs. \NAME therefore implements \emph{stream time shifting}: it delays the client-visible emission schedule while leaving the GPU execution schedule untouched.

Let $a_i$ be the gateway arrival time of the $i$-th streamed chunk from the inference engine, measured on a monotonic clock, and let $t_0$ be request arrival time. The physical TTFT is $t_{\mathsf{actual}}=a_1-t_0$. In Phase 2, \NAME samples a target first-token latency $\tau$ from the calibrated miss distribution and computes
\begin{equation}
\Delta_{\mathsf{pad}}=\max(0,\tau-t_{\mathsf{actual}}).
\end{equation}
The emitted time of every chunk is
\begin{equation}
e_i=a_i+\Delta_{\mathsf{pad}}, \qquad i=1,2,\ldots
\label{eq:timeshift}
\end{equation}
so the observed first-token latency is $e_1-t_0=\max(t_{\mathsf{actual}},\tau)$, while every inter-token gap is preserved:
\begin{equation}
e_i-e_{i-1}=a_i-a_{i-1}, \qquad i\ge 2.
\end{equation}

\NAME realizes Equation~\ref{eq:timeshift} inside the tokenizer manager. When a request is admitted, \NAME records a monotonic request timestamp in the per-request state. Immediately before each streamed output chunk reaches the HTTP emission path, \NAME stamps the chunk with a GPU-side arrival timestamp.
On the first chunk, the gateway computes $t_{\mathsf{actual}}$, reads prompt-token and cache-hit metadata from the scheduler, consults the Privacy Odometer, records the observation for calibration, and stores $\Delta_{\mathsf{pad}}$ in the request state. On every later chunk, the gateway releases the chunk at its stamped arrival time plus the same stored offset. 

\noindent\textbf{From per-request to per-batch equalization.}
The time shift above equalizes the emission surface at the level of a \emph{single} request, making its first-token latency miss-like irrespective of the cache bit. A shared backend exposes one further observable: when an attacker co-submits several probes under the same prefix, a longest-prefix-match scheduler serves them in prefix-priority order, so the \emph{order} in which their responses return reflects which candidate matched the victim's cache. \NAME closes this cross-request channel under the \emph{same} phase gate that governs the time shift.

The gateway groups the concurrent responses of each locked-down prefix, identifying a group by the metered prefix hash, the same anchor $P$ the odometer scores; all Phase-2 probes of one protected prefix therefore land in one group, from whatever account or connection they were submitted. Within a group, responses are held on an adaptive barrier that releases once the group has been quiet for a short settle interval, with an upper bound so that a stalled response cannot hold the group. At release, every response draws an independent uniform offset, and independent draws induce a uniformly random emission permutation: the order carries no information about which candidate hit. Two properties make this step cheap and safe. First, the barrier only ever \emph{adds} delay on top of the per-request clamp: a grouped response is emitted at the later of its own miss target $\tau$ and the group's release time, so the magnitude guarantee of Lemma~\ref{lem:advantage}, that every emitted first token sits at or above the miss tail, is untouched; the batch step removes the order signal without reopening the magnitude one. Second, the hold is confined to attack traffic: benign requests (Phase~0/1) are never grouped or reordered, and the responses that do wait belong to a prefix already under active fan-out and already being padded toward the same miss target. An attacker can escape the group only by spacing its probes out, which forfeits the co-submitted ordering the channel needs; each isolated probe then faces the per-request time shift alone. Appendix~\ref{app:mrr} confirms the mechanism still closes the order channel under real batched load ($\text{MRR}{=}16$), with the magnitude attacks unaffected.

\noindent\textbf{Choosing the emission target.}
\label{sec:method-cdf}
It remains to choose the target latency $\tau$ in Equation~\ref{eq:timeshift}. Because the whole-stream shift already erases every inter-token cue (Lemma~\ref{lem:advantage}), only the emitted first-token law matters. \NAME's emitter is a one-sided clamp to a calibrated upper miss quantile $Q$,
\begin{equation}
e_1-t_0=\max(t_{\mathsf{actual}},\,Q),\qquad Q=Q_q\big(\widehat F_{\mathsf{miss}}\big),
\label{eq:clamp}
\end{equation}
where $Q$ is the $(1+\eta)$-scaled $q$-th quantile ($q=0.97$) of the online miss distribution $\widehat F_{\mathsf{miss}}$. \NAME applies this clamp to all requests under a Phase-2 prefix, regardless of their actual cache outcome. Consequently, physical hits and approximately $97\%$ of genuine misses collapse onto the exact same latency floor $Q$. Because no unpadded baseline exists under a locked prefix, hit and miss candidates become statistically indistinguishable, rendering multi-probe statistical averaging futile (validated by the adaptive distinguisher in Section~\ref{sec:adaptive}). 

Formally, Lemma~\ref{lem:advantage} bounds the attacker's residual advantage by the equalization gap $\delta_{\mathsf{eq}}$. Although clamping to the $97$-th percentile incurs higher latency than padding to the mean miss latency, it is necessary: as shown in Section~\ref{sec:ablation}, lower-overhead distribution-matching emitters fail to reliably mask tail discrepancies and leave exploitable residuals. Clamping thus provides robust, worst-case indistinguishability while remaining substantially faster than cache isolation (Section~\ref{sec:performance}).

\subsection{Dynamic Miss Estimator}
\label{sec:estimator}
To ensure robust indistinguishability, the target padding latency $\tau$ must dynamically track the true cache-miss distribution under fluctuating server loads, batch sizes, and queueing delays. Static calibration is insufficient, as it risks under-padding during load spikes or adding excessive latency when the server is idle. \NAME resolves this by coupling an affine latency model with online regression.

\noindent\textbf{Affine Latency Modeling.}
Cold-miss TTFT does not scale purely proportionally with prompt length $L$. Instead, it exhibits an affine relationship:
\begin{equation}
\mathrm{TTFT} \approx a + bL,
\end{equation}
where the intercept $a$ captures constant system overheads (e.g., tokenization, scheduling, and kernel launch latency), while the slope $b$ reflects the marginal prefill cost per token. Because a strictly proportional model severely misestimates the execution time across varied prompt lengths, \NAME estimates latency directly in raw TTFT space.

\noindent\textbf{Online Estimation and Target Generation.}
For each request that experiences a full cache miss ($\texttt{cached\_tokens}=0$), \NAME logs the observed $(L, \mathrm{TTFT})$ tuple into a sliding window. It continuously updates the parameters $(a, b)$ via streaming least-squares regression and computes the residuals $R = \mathrm{TTFT} - (a + bL)$. The upper quantile of these residuals, $Q_q(R)$, captures the real-time tail behavior of cache misses. For a protected request of length $L$, the target latency $\tau$ is computed as:
\begin{equation}
\tau = \max \Big\{ (a + bL)(1+\eta),\ \big(a + bL + Q_q(R)\big)(1+\eta) \Big\} + \epsilon,
\label{eq:tau-affine}
\end{equation}
where $\eta$ is a safety margin, $q$ is the target quantile, and $\epsilon \sim \mathcal{U}(0, \sigma)$ injects bounded random jitter to eliminate deterministic timing signatures.

\noindent\textbf{Cold-Start and Fallback Mechanisms.}
To guarantee protection when the regression window is still filling or contains degenerate prompt lengths, \NAME maintains secondary Exponential Moving Average (EMA) tracks for latency:
\begin{equation}
\mu_z(t) = \beta \mu_z(t-1) + (1-\beta)x_t, \quad z \in \{\mathsf{hit}, \mathsf{miss}\}.
\label{eq:ema}
\end{equation}
Pure misses feed the miss track, while partial hits update the hit track after subtracting the estimated cost of their uncached suffix. During initial system warmup, \NAME adopts conservative default bounds, ensuring that requests remain fully protected from the moment the server boots.

\section{Security and Performance Analysis}
\label{sec:security}
\subsection{The Importance of the Fan-out Signal ($\phi$)}
\NAME's threshold is meaningful precisely because a timing attack must spend many candidate probes, and therefore open many novel fan-out branches under the same prefix, to resolve a secret. We formalize this fan-out requirement with a standard information-theoretic lower bound.

\begin{theorem}[Fan-out lower bound]\label{thm:fanout}
Let $S^*$ be the unknown $N$-token secret continuation after prefix $P$. Let $\mathbf{O}=(O_1,\ldots,O_k)$ be the timing observations from $k$ adaptive oracle queries. If each query has channel capacity at most $C$ bits, then any attacker with recovery error at most $\varepsilon$ must satisfy
\begin{equation}
k \ge
\frac{H(S^*|P)-h_2(\varepsilon)-\varepsilon\log_2(M-1)}{C},
\label{eq:fano}
\end{equation}
where $M=|\mathrm{supp}(S^*)|$ is the support size of $S^*$, i.e., the number of distinct candidates the secret continuation could take and the attacker must disambiguate, and $h_2$ is binary entropy. As $\varepsilon \rightarrow 0$, the lower bound tends to $H(S^*|P)/C$.
\end{theorem}

\begin{proof}
By the chain rule for mutual information, adaptivity does not let the attacker exceed the per-query channel capacity:
\begin{equation}
I(S^*;\mathbf{O}|P)\le \sum_{i=1}^{k} I(S^*;O_i|P,O_{<i}) \le kC.
\label{eq:chainrule}
\end{equation}
Fano's inequality gives
\begin{equation}
H(S^*|\mathbf{O},P)\le h_2(\varepsilon)+\varepsilon\log_2(M-1).
\end{equation}
Since $H(S^*|\mathbf{O},P)=H(S^*|P)-I(S^*;\mathbf{O}|P)$, rearranging yields Equation~\ref{eq:fano}. The limiting statement follows because both $h_2(\varepsilon)$ and $\varepsilon\log_2(M-1)$ go to zero for fixed finite $M$.
\end{proof}
How large is the per-query capacity $C$? The binary hit/miss verdict alone gives $C\le1$ bit, but the observable is a \emph{continuous} TTFT whose magnitude also encodes how much of the prompt was cached. Since identifying the exact match depth among $L$ possible token positions resolves $\log_2 L$ bits of uncertainty, an \emph{unpadded} probe can carry up to $O(\log L)$ bits about the matched-prefix length.
\NAME's two phases price this honestly. In the grace phase, probes are unpadded, so Theorem~\ref{thm:fanout} must be read with this larger $C$, which is exactly why the grace budget is kept small ($G{=}1$). In lockdown, the clamp pushes every emitted first token to the miss tail, erasing the magnitude together with the hit/miss bit, so the per-probe capacity collapses to the equalization gap $\deq$ bounded below (Lemma~\ref{lem:advantage}). Theorem~\ref{thm:composition} then composes the two phases: a few grace probes at full capacity, and every further probe at only $\deq$.

A complementary probability view turns the bound into a parameter rule. Let
\begin{equation}
p_B=\max_s \Pr[S^*=s|P]
\end{equation}
denote the attacker's maximum prior probability of correctly guessing the exact protected secret ($S^*$) given only the known context prefix ($P$). In essence, $p_B$ quantifies the attacker's best-case baseline success rate \emph{before} observing any timing side-channel signals, leveraging their strongest language model or background knowledge to predict the most likely continuation $s$. 
Let $H_{\infty}(S^*|P)=-\log_2 p_B$ be the corresponding conditional min-entropy, which formally quantifies this fundamental unpredictability of the secret. Since \NAME allows an attacker to evaluate at most $K_{\max}(P)$ distinct Phase-1 candidates before triggering lockdown, a union bound yields the maximum probability of extracting the secret without being penalized:
\begin{equation}
\begin{aligned}
\Pr[\mathsf{Phase\ 1\ success}]
&\le \min\{1,K_{\max}(P)p_B\}\\
&= \min\{1,K_{\max}(P)2^{-H_{\infty}(S^*|P)}\}.
\end{aligned}
\label{eq:union}
\end{equation}

Therefore, a deployment with a strictly defined target Phase-1 leakage probability $\alpha$ should configure its grace budget as:
\begin{equation}
G=K_{\max}(P)\le \left\lfloor \alpha\,2^{H_{\infty}(S^*|P)} \right\rfloor
\end{equation}
and subsequently set $G\,\omega_{\mathsf{sens}}\le\theta<(G+1)\omega_{\mathsf{sens}}$. Our primary configuration ($\theta = 1.0, \omega_{\mathsf{sens}} = 1.0$) is the case $G=1$: at most one unpadded candidate is admitted, so $\Pr[\mathsf{Phase\ 1\ success}]\le p_B$, the prior guessing probability. After lockdown (Phase 2), the emitted timing distribution is forcefully equalized to the calibrated miss distribution. Consequently, the protected hit/miss bit contributes zero additional timing advantage in the ideal equalization model, reducing any residual success purely to the attacker's non-timing prior guessing ($p_B$).

\subsection{The Security of the Time-Decay Factor ($\gamma$)} The time-decay factor $\gamma$ is not intended to let an attacker probe forever at a low rate. KV cache entries are volatile data-plane state. Under normal multi-tenant load, cache eviction can be modeled as a stochastic process with rate $\lambda$. If evictions of the target prefix are approximated by a Poisson process, the probability that the target cache remains alive for duration $T$ is
\begin{equation}
\Pr[\mathsf{survive}\ T]=e^{-\lambda T}.
\end{equation}
An attacker who needs $k$ candidate probes and waits average interval $\Delta t$ spends approximately $T=k\Delta t$ time, giving
\begin{equation}
\Pr[\mathsf{cache\ survives\ attack}]\approx e^{-\lambda k\Delta t}.
\label{eq:survive}
\end{equation}
Large $\Delta t$ helps the odometer decay, but it exponentially reduces the chance that the victim's cache state survives long enough to be queried. Small $\Delta t$ preserves cache survival, but then $\gamma^{\Delta t}\approx1$ and the score in Equation~\ref{eq:score} quickly crosses $\theta$. This creates the adversary's spatio-temporal dilemma: to keep the cache alive, the attacker must probe quickly; probing quickly exposes the high-fan-out pattern that \NAME meters.

This also gives a direct rule for choosing $\gamma$. If a prefix has current score $S_0>\theta$, the idle time needed to decay back below the grace boundary is
\begin{equation}
T_{\mathsf{decay}}(S_0,\theta,\gamma)
=\frac{\ln(\theta/S_0)}{\ln\gamma}.
\label{eq:decay-time}
\end{equation}
For a prefix that has just entered lockdown at $S_0=2$ (two rapid novel probes), our default $\theta=1.0$ and $\gamma=0.995$/s require about $138$\,s of idle time before it decays back below the grace boundary. Thus $\gamma$ should be selected so that $T_{\mathsf{decay}}$ is longer than the time window over which a provider expects a sensitive KV entry to remain reliably resident under normal load, or longer than the cadence at which repeated probes can be rate-limited. Decreasing $\gamma$ shortens memory and improves benign liveness; increasing $\gamma$ lengthens lockdown and closes low-and-slow repetition more aggressively.

\subsection{Conditional Equalization Guarantee}
Theorem~\ref{thm:fanout} shows that resolving a secret needs many novel
probes; \NAME admits those probes only once the prefix is in lockdown, where
their timing is equalized. We state that guarantee conditionally and then
compose it with Theorem~\ref{thm:fanout}.

Fix a metered prefix $P$ and let $B \in \{\hit,\miss\}$ indicate whether a
probe's continuation matches the victim's cached prefix, under an
\emph{arbitrary} prior $\pi = \Pr[B = \hit]$. The attacker observes the whole
emitted stream
\begin{equation*}
  \Tr = (o_1, g_2, \dots, g_n), \quad
  o_1 = e_1 - t_0, \quad g_i = e_i - e_{i-1},
\end{equation*}
i.e.\ the first-token latency and every inter-token gap. Write $\Lo^{o}_b$
and $\Lo^{\Tr}_b$ for the laws of $o_1$ and of $\Tr$ given $B = b$.

\begin{assumption}[Equalization regime]\label{as:regime}
\emph{(A1) Gating and lockdown:} $\omega_{\mathsf{sens}}(P) > 0$, every
distinct candidate probe maps to the same metered prefix $P$, and $P$ is in
Phase~2, i.e.\ $S_P(t) > \theta$.
\emph{(A2) Decode-schedule conditional independence:} there is a Markov
kernel $K$ with $\Lo(g_{2:n} \mid B = b,\, o_1 = o) = K(\cdot \mid o)$ for
a.e.\ $o$ and both $b \in \{\hit,\miss\}$.
\emph{(A3) First-token equalization:}
$\TV(\Lo^{o}_{\hit}, \Lo^{o}_{\miss}) \le \deq$.
\end{assumption}

(A2) is where the whole-stream claim is earned. \NAME applies a \emph{single}
offset $\Delta_{\mathsf{pad}}$ per chunk, so by \eqref{eq:timeshift} every gap
satisfies $e_i - e_{i-1} = a_i - a_{i-1}$: the emitter never manufactures a
burst. Gaps are not \emph{unconditionally} independent of $B$ (decode cadence
and prefill are load-driven); (A2) asks only that \emph{given $o_1$}, $B$ adds
nothing further. The dynamic miss estimator (Section~\ref{sec:estimator}) buys
exactly this: $\tau$ is resampled from the \emph{current} miss window, so a
padded hit and a natural miss sharing $o_1$ share the load regime that produced
their gaps. It is falsifiable: the distinguisher of
Section~\ref{sec:adaptive} tests it directly.

\begin{lemma}[Trace reduction]\label{lem:reduction}
Under (A2), $\TV(\Lo^{\Tr}_{\hit}, \Lo^{\Tr}_{\miss})
= \TV(\Lo^{o}_{\hit}, \Lo^{o}_{\miss})$.
\end{lemma}

\begin{proof}
See Appendix~\ref{app:proofs}.
\end{proof}

\begin{lemma}[Bounded advantage and capacity]\label{lem:advantage}
Under Assumption~\ref{as:regime}, for a single probe: \emph{(a)} every
(possibly randomized) test on the whole trace $\Tr$ has advantage at most
$\deq$ in deciding $B$; \emph{(b)} for \emph{every} prior $\pi$,
$I(B;\Tr) \le h_2(\pi)\,\deq \le \deq$ bits, so the per-query capacity
satisfies $C_{\mathsf{eq}} := \max_\pi I(B;\Tr) \le \deq$. As $\deq \to 0$
both vanish.
\end{lemma}

\begin{proof}
See Appendix~\ref{app:proofs}.
\end{proof}

Part (b) says: with probability $1-\deq$ the emitted latency is drawn from a
law common to hits and misses, hence uninformative; only with probability
$\deq$ can it betray the bit. It subsumes the uniform-prior statement
$I(B;o_1)=\mathrm{JSD}\le\deq$ ($\pi=\tfrac12$, $h_2=1$) and, unlike it, holds
for \emph{every} prior, which is exactly what Theorem~\ref{thm:composition} needs.

\begin{corollary}[Repeated measurement]\label{cor:repeat}
If the attacker repeats a probe $R$ times with traces conditionally i.i.d.\
given $B$, its advantage is at most $1-(1-\deq)^R$ and
$I(B;\Tr^{(R)}) \le h_2(\pi)\bigl[1-(1-\deq)^R\bigr]$.
\end{corollary}

\begin{proof}
See Appendix~\ref{app:proofs}.
\end{proof}

\begin{theorem}[Odometer--equalization composition]\label{thm:composition}
Let an attacker issue $k$ probes under $P$, of which $N_1$ are admitted in
Phase~1, with Assumption~\ref{as:regime} holding for every Phase-2 probe.
Then recovering $S^*$ with error at most $\varepsilon$ requires
\begin{equation}
  k \;\ge\; N_1 \;+\;
  \frac{H(S^*\!\mid\! P) - h_2(\varepsilon) - \varepsilon \log_2(M-1) - N_1}
       {\deq}.
  \label{eq:composition}
\end{equation}
\end{theorem}

\begin{proof}
See Appendix~\ref{app:proofs}.
\end{proof}

Equation~\eqref{eq:composition} \emph{refines} rather than contradicts \eqref{eq:fano}: setting $N_1 = 0$ and $\deq = C$ recovers Theorem~\ref{thm:fanout} exactly, and since $\deq \le 1$, the composed bound is never weaker. The fundamental difference lies in the information accounting. While Theorem~\ref{thm:fanout} assumes every probe leaks the same maximum capacity, Theorem~\ref{thm:composition} models \NAME's two-phase reality: it charges the initial $N_1$ grace probes at full capacity, but restricts every subsequent probe to a minimal leakage of $\deq$.

Crucially, this mathematical composition mirrors how \NAME's three physical mechanisms collaborate to corner the attacker: First, the \emph{odometer} strictly caps $N_1$, allowing at most $K_{\max}(P)$ unpadded probes per decay window (Equation~\ref{eq:kmax}). Second, \emph{equalization} drastically inflates the attack cost during lockdown, forcing the adversary to spend $1/\deq$ probes to extract just one additional bit. Finally, natural \emph{cache residency} (Equation~\ref{eq:survive}) closes the low-and-slow evasion loophole, physically bounding how many decay windows an attacker can patiently wait out before the targeted cache entry is evicted.

The guarantee is conditional: it degrades if the gate misfires or probes split
across metered prefixes (A1), if the schedule leaks beyond $o_1$ (A2), or if
$\deq > 0$ (A3); Appendix~\ref{sec:perf-bounds} in turn bounds this gap $\deq$
by the tunable clamp knobs $(q,\eta)$, pricing security directly in padding
latency. Finally, $\deq$ is a \emph{measured} quantity, not a bound
over all conceivable tests: any concrete distinguisher only lower-bounds the
true total variation. Because the optimal test on a balanced population
attains accuracy $\tfrac12(1+\TV)$, a distinguisher's held-out
\emph{balanced accuracy} $\mathrm{acc}$ (the mean of its true-positive and
true-negative rates) yields the estimate
$\widehat{\delta}_{\mathsf{eq}} = 2\,\mathrm{acc} - 1$, which grows as the
distinguisher strengthens; a residual that survived a stronger future test
would raise $\widehat{\delta}_{\mathsf{eq}}$ and, through
\eqref{eq:composition}, the number of probes that it implies.
Finally, Appendix~\ref{sec:perf-bounds} shows that benign traffic remains unpadded and GPU demand is zero. Proposition~\ref{prop:dial} bounds $\deq \le (1-q)+\varepsilon_{\mathsf{est}}+p_{\mathsf{slow}}$, directly linking the deployment knobs $(q,\eta)$ to the empirically measured residual $\widehat{\delta}_{\mathsf{eq}}$.

\section{Experiments}
\label{sec:experiments}
We integrate \NAME into SGLang~\cite{zheng2024sglang}, a state-of-the-art serving framework featuring a native radix-tree prefix cache. Our evaluation answers three questions: (Q1) how effectively does \NAME suppress timing and response-order side channels relative to these baselines; (Q2) what serving-performance cost does it impose on benign public-prefix traffic; and (Q3) does it remain robust against an adaptive adversary with full knowledge of the mechanism (Section~\ref{sec:adaptive})?

\subsection{Experimental Setup}
\noindent Our full experimental setup, the hardware and serving configuration, the models and datasets, and \NAME's defense configuration, is deferred to Appendix~\ref{app:setup}; we summarize the attacks, baselines, and metrics here.

\noindent\textbf{Attacks and baselines.} We evaluate three KV-cache timing attacks that span the leakage spectrum. \emph{EarlyBird}~\cite{early-bird} recovers a single next token from per-request TTFT magnitude. \emph{InputSnatch}~\cite{input-snatch} recovers a structured (name, condition) field from the same magnitude signal; a matching field differs by only a few prefix tokens. \emph{OptiLeak}~\cite{wang2026optileak} exploits the response \emph{order} of a co-submitted probe batch under SGLang's longest-prefix-match scheduling, a strong order channel. We compare against four points: an undefended \textit{No Defense}, \textit{Cache Partition}~\cite{cache-partitioning} (per-tenant radix-tree namespaces), \textit{CacheSalt}~\cite{vllmCacheSalt} (per-request salts/extra cache keys), and \textit{SafeKV}~\cite{safeKV} (a sensitivity-aware proxy that isolates prompts classified as sensitive).

\noindent\textbf{Metrics.}
\phantomsection\label{sec:metrics}
We quantify attack efficacy using $\mathrm{ASR}@N$, defined as the attack success rate within a budget of $N$ candidate queries. An evaluation unit represents one target to recover: a hidden private entity for EarlyBird and InputSnatch, and a sensitive token position for OptiLeak. Specifically, EarlyBird searches over the full candidate pool of $467$ first-tokens, while InputSnatch performs end-to-end recovery over the complete $500{\times}500$ (name, condition) dictionary without artificial ground-truth injection. For OptiLeak, candidates are evaluated in co-submitted batches of $\le\!50$ along the model's prediction ranking up to rank $10{,}000$. 
We report $95\%$ percentile bootstrap confidence intervals for every headline number; the resampling methodology and per-cell intervals are in Appendix~\ref{app:asr-ci}.

\subsection{Security Effectiveness}

Table~\ref{tab:asr} reports the ASR@N across the three evaluated models. Without defenses, all three timing channels leak significantly. This vulnerability worsens with model scale, as larger models exhibit longer prefill times that create a sharper latency gap between cache hits and misses. For example, OptiLeak's order channel achieves an ASR@10k of $67.6$--$81.8\%$. EarlyBird's magnitude channel reaches $92.4\%$ on DeepSeek-14B, dropping to $6.2\%$ on the smaller Qwen1.5-1.8B. InputSnatch, which recovers a complex multi-token field, follows the exact same trend. When evaluated over its full dictionary to avoid artificially masking the attack, its end-to-end recovery climbs from the $0.0\%$ guessing floor on Qwen (where the minimal TTFT gap is lost in system noise) all the way to $59.4\%$ on the DeepSeek-14B model.

\NAME mitigates all three attacks using a single, unmodified policy. Across all models, OptiLeak's success rate drops to exactly $0.0\%$, while EarlyBird falls to near-zero ($0.0$--$0.2\%$). InputSnatch is similarly suppressed to $\le\!0.4\%$ end-to-end. Although a negligible residual timing signal survives for individual tokens on larger models due to the equalization gap, the combined fields remain strictly at chance, ensuring that full end-to-end extraction consistently fails.

Baseline defenses like Cache Partition and CacheSalt also successfully suppress these channels to the random guessing floor ($0.0$--$1.0\%$ across all attacks and budgets). However, they achieve this security exclusively by disabling cross-tenant cache reuse entirely, forcing redundant model recomputation. SafeKV, which instead shares selectively, leaves a slightly larger residual (up to $2.0\%$ EarlyBird on Qwen1.5-1.8B) wherever its classifier misjudges a sensitive prefix as public. Ultimately, \NAME achieves the same near-zero ASR as these isolation defenses, but crucially does so while preserving the computational benefits of cache sharing for benign traffic.
\begin{table*}[t]
\centering
\caption{Attack Success Rate ($\mathrm{ASR}@N$, \%) across model scales and defenses.}\label{tab:asr}
\scriptsize
\setlength{\tabcolsep}{3.5pt}
\resizebox{0.68\textwidth}{!}{
\begin{tabular}{ll rrr r rrr}
\toprule
\multirow{2}{*}{\textbf{Model}} & \multirow{2}{*}{\textbf{Defense}} &
\multicolumn{3}{c}{\textbf{EarlyBird} $\downarrow$} &
\multicolumn{1}{c}{\textbf{InputSnatch} $\downarrow$} &
\multicolumn{3}{c}{\textbf{OptiLeak} $\downarrow$} \\
\cmidrule(lr){3-5}\cmidrule(lr){6-6}\cmidrule(lr){7-9}
& & \textbf{ASR@100} & \textbf{ASR@1k} & \textbf{ASR@10k} & \textbf{ASR@100--10k} & \textbf{ASR@100} & \textbf{ASR@1k} & \textbf{ASR@10k} \\
\midrule
\multirow{5}{*}{DeepSeek-14B}
 & No Defense      & 21.6 & 92.4 & 92.4 & 59.4 & 25.5 & 41.2 & 68.8 \\
 & Cache Partition & 0.0 & 0.5 & 0.5 & 0.0 & 0.0 & 0.0 & 0.0 \\
 & CacheSalt       & 0.0 & 0.0 & 0.0 & 0.0 & 0.0 & 0.0 & 0.0 \\
 & SafeKV          & 0.0 & 0.5 & 0.5 & 0.0 & 0.0 & 0.0 & 0.0 \\
 & \textbf{\NAME}  & \textbf{0.0} & \textbf{0.2} & \textbf{0.2} & \textbf{0.4} & \textbf{0.0} & \textbf{0.0} & \textbf{0.0} \\
\midrule
\multirow{5}{*}{Llama-3-8B}
 & No Defense      & 8.2 & 33.4 & 33.4 & 34.2 & 29.2 & 48.8 & 81.8 \\
 & Cache Partition & 0.0 & 1.0 & 1.0 & 0.0 & 0.0  & 0.0  & 0.0 \\
 & CacheSalt       & 0.0 & 0.0 & 0.0 & 0.0 & 0.0  & 0.0  & 0.0 \\
 & SafeKV          & 0.0 & 0.5 & 0.5 & 0.0 & 0.0  & 0.0  & 0.0 \\
 & \textbf{\NAME}  & \textbf{0.0} & \textbf{0.0} & \textbf{0.0} & \textbf{0.4} & \textbf{0.0} & \textbf{0.0} & \textbf{0.0} \\
\midrule
\multirow{5}{*}{Qwen1.5-1.8B}
 & No Defense      & 2.2 & 6.2 & 6.2 & 0.0 & 24.1 & 38.9 & 67.6 \\
 & Cache Partition & 0.0 & 0.5 & 0.5 & 0.0 & 0.0  & 0.0  & 0.0 \\
 & CacheSalt       & 0.0 & 0.0 & 0.0 & 0.0 & 0.0  & 0.0  & 0.0 \\
 & SafeKV          & 0.0 & 2.0 & 2.0 & 0.0 & 0.0  & 0.0  & 0.0 \\
 & \textbf{\NAME}  & \textbf{0.0} & \textbf{0.0} & \textbf{0.0} & \textbf{0.0} & \textbf{0.0} & \textbf{0.0} & \textbf{0.0} \\
\bottomrule
\end{tabular}}
\end{table*}

\subsection{Performance Overhead}
\label{sec:performance}

\begin{figure*}[t]
\centering
\includegraphics[width=0.7\textwidth]{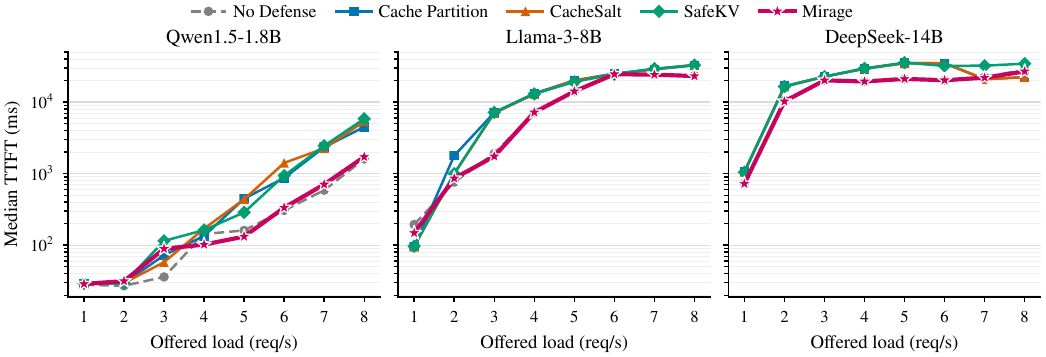}
\caption{Median TTFT (log scale) vs.\ offered load on public traffic. \NAME (pink) closely tracks No~Defense (grey dashed, largely occluded); isolation baselines sit above. Curves converge only in the saturated tail, where queueing dominates.}
\Description{Three line plots (Qwen1.5-1.8B, Llama-3-8B, DeepSeek-14B) showing median TTFT versus offered load for No Defense, Cache Partition, CacheSalt, SafeKV, and Mirage. Mirage stays close to No Defense while the isolation baselines are higher.}
\label{fig:ttft-load}
\end{figure*}

We evaluate serving overhead using public Alpaca prompts generated by 500 virtual users under an open-loop Poisson arrival process. We sweep the offered load from 1 to 8 requests per second, recording the median TTFT and goodput at each point (Figure~\ref{fig:ttft-load}). To cleanly isolate the defense's inherent overhead from severe system queueing delays, all single-value latency metrics reported below are measured at each model's peak unsaturated load. This is defined as the maximum rate where goodput maintains at least $90\%$ of the offered load ($1$, $2$, and $6$ req/s for the DeepSeek-14B, Llama-3-8B, and Qwen1.5-1.8B models, respectively). 

For public traffic, \NAME imposes negligible overhead. Its median TTFT remains within $1.5\%$ of the undefended baseline ($55.2$\,ms vs.\ $54.4$\,ms in a controlled $300$-request trial). In contrast, strict isolation mechanisms like Cache Partition and CacheSalt penalize all traffic indiscriminately, inflating DeepSeek-14B's median TTFT by $44$--$46\%$ and causing multi-fold latency increases on Qwen1.5-1.8B at higher loads. SafeKV, a selective isolation system, matches \NAME and the undefended baseline here, provided its detection runs off the critical path and correctly leaves public traffic untouched. Ultimately, on purely public workloads, \NAME matches the efficiency of ideally configured selective sharing while thoroughly outperforming unconditional isolation. This advantage widens sharply under contended sensitive-shared workloads: when the GPU is saturated, isolation's redundant recomputation inflates TTFT by roughly $3\times$ while \NAME reuses the cached prefix and tracks No~Defense (Appendix~\ref{sec:sensshared}).

Having established \NAME's efficiency on public workloads, we now evaluate the precision of its behavioral trigger on sensitive traffic.
To demonstrate that this padding is applied exclusively when necessary, we isolate the trigger mechanism by comparing \NAME at its default grace budget ($G{=}1$) against two absolute bounds: No Defense (always off) and a Static Pad configuration ($G{=}0$, where every sensitive request is unconditionally padded).

We measure the median TTFT of a sensitive cache access under two conditions: a single legitimate query (benign first-visit) and the exact same query issued during an active fan-out attack. By keeping the baseline request rate low, we allow the odometer's time-decay to naturally spare genuine traffic while rapidly locking down deliberate probing bursts. As shown in Table~\ref{tab:phasemix}, No Defense processes both conditions at the native $\approx\!150$\,ms speed but leaves the timing channel wide open (ASR up to $92\%$, Table~\ref{tab:asr}). Conversely, the Static Pad configuration securely closes the channel but severely penalizes the user experience, inflicting a $417$\,ms delay even on the benign first-visit. \NAME effortlessly bridges this gap. It serves the benign user at the pristine $150$\,ms latency (recording zero Phase-2 events) while dynamically applying the $418$\,ms obfuscation solely when the prefix is under attack.

\begin{table}[t]
\centering
\caption{Median TTFT (ms) of a sensitive access (Llama-3-8B): \NAME pads only under attack; pad-all pads every sensitive request.}
\label{tab:phasemix}
\resizebox{0.78\columnwidth}{!}{
\begin{tabular}{l cc c}
\toprule
\textbf{Configuration} & \textbf{Benign first-visit} & \textbf{Under attack} & \textbf{Attack} \\
 & \textbf{TTFT (ms)} & \textbf{TTFT (ms)} & \textbf{ASR} \\
\midrule
No Defense              & $149$ & $151$ & leaks \\
\textbf{\NAME ($G{=}1$)} & $\mathbf{150}$ & $418$ & $\approx0$ \\
Naive pad-all ($G{=}0$) & $417$ & $416$ & $\approx0$ \\
\bottomrule
\end{tabular}}
\end{table}

\begin{wrapfigure}{r}{0.55\columnwidth}
  \centering
  \includegraphics[width=\linewidth]{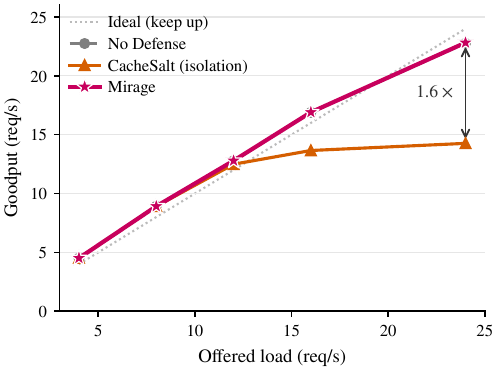}
  \caption{Goodput vs.\ offered load (Llama-3-8B). \NAME tracks No~Defense along the ideal line; per-tenant isolation saturates $\sim\!1.6\times$ earlier.}
  \label{fig:goodput}
\end{wrapfigure}

\noindent\textbf{Throughput headroom.} The latency sweep above uses one in-flight request ($\text{MRR}{=}1$), the adversary-favorable setting that sharpens the timing signal. We now set $\text{MRR}{=}16$ to run the GPU as a real batched server and stress throughput. \NAME spends no GPU cycles on the defense (the pad is at the gateway and the shared prefix is reused), whereas isolation recomputes the shared prefix for every tenant, burning cycles that could serve new requests. On Llama-3-8B (Figure~\ref{fig:goodput}), \NAME's goodput tracks No~Defense along the ideal line while per-tenant isolation saturates near $14$~req/s; at $24$~req/s \NAME sustains about $1.6\times$ its throughput, matching the capacity ratio that Theorem~\ref{thm:capacity} predicts. Because the pad never stalls the GPU, padding the sensitive share of this traffic ($5\%$ here) leaves \NAME's goodput indistinguishable from No~Defense. This headroom, distinct from the latency result of Figure~\ref{fig:ttft-load}, is what isolation forfeits and \NAME retains.

\noindent\textbf{Memory and control-plane cost.} The Privacy Odometer lives entirely in the host tokenizer manager and adds \emph{zero} GPU memory. Each tracked prefix costs about $520$\,B plus $\approx100$\,B per distinct branch, so tracking $10{,}000$ active sensitive prefixes uses $\approx35$\,MiB, negligible beside the weights and KV cache. On the critical path, a full odometer decision (sensitivity scan, novelty hash, decayed-score update, phase gate) costs a median $\approx6.4\,\mu$s per request, roughly four orders of magnitude below a typical $\approx150$\,ms TTFT; this is the constant control-plane term $c_{\mathsf{ctrl}}$ that Theorem~\ref{thm:latency} proves is \NAME's only latency cost on benign traffic.

\subsection{Generalization to Semantic Caches}
\label{sec:semantic}
We now test whether the defense generalizes beyond KV/prefix caches to a
\emph{semantic cache}. A GPTCache-style semantic cache embeds each query and
returns the stored response of a sufficiently similar earlier query, so a hit
answers in milliseconds while a miss runs the full model. The hit/miss gap is
therefore a whole response ($70$--$450$\,ms in our runs), much larger than the
per-token prefill gap of Section~\ref{sec:experiments}, which makes the
undefended channel wide open and the layer a harder test for the defense.

We lift the same two-phase design to this layer. Only the novelty signal has to change, because an embedding
space has no exact continuation hash to key on. Consider first how the attacker behaves. To pull a secret out of a semantic cache, an attacker has to work
a small neighborhood of the embedding space hard: it fires a rapid burst of
queries that are all worded almost the same, sweeping the candidate answers one
after another. A benign user does the opposite: they ask about the one entity they
care about, rephrasing it at most a few times. This gives two
behavioral tells that a benign user does not produce together, a sudden
\emph{density} of near-duplicate queries in one region, and a sweep \emph{across
distinct entities}. The separation is wide in practice: in our corpus, benign
paraphrases of one topic have an average pairwise cosine of $0.835$, while an
attacker's probes of different entities sit at $0.445$. 

\NAME-Semantic meters exactly these two tells with two matching components,
\emph{semantic-differential hashing} for the entity sweep and \emph{local
density tracking} for the burst. The detector maintains a set of \emph{entity representatives}. If a new query's similarity to an existing representative exceeds a threshold ($s_e{=}0.80$), it collapses onto that entity; otherwise, it registers as a novel entity. We then track the time-decayed density of distinct entities queried within a region:
\begin{equation}
D_{\mathsf{ent}}(t_j)=\gamma^{\,t_j-t_{j-1}}D_{\mathsf{ent}}(t_{j-1})+\phi_{\mathsf{ent}}(q_j),
\label{eq:sem-density}
\end{equation}
where $\phi_{\mathsf{ent}}(q_j)=1$ only when a query opens a new entity. Benign paraphrases map to the same representative and never raise the risk score. However, an attacker's rapid sweep across distinct entities quickly pushes $D_{\mathsf{ent}}$ past a safety threshold ($\gamma{=}0.98$/s, floor $4$). Once triggered, the region enters lockdown, and hit latencies are padded up to the miss distribution.

We evaluate this mechanism on a $40$-topic medical corpus using an \repo{all-MiniLM-L6-v2} embedder and the three backend models from Table~\ref{tab:asr}. Because standard KV-cache attacks do not apply to semantic caches, we port their underlying objectives: (i) \textit{membership} (detecting whether a target topic was queried), (ii) \textit{closed-world} (recovering the exact topic among $K{=}10$ candidates), and (iii) \textit{open-world} (recovering the topic among all $K{=}40$ candidates). We run $100$ trials per objective. Since these are classification tasks, we evaluate success against random guessing baselines of $50\%$, $10\%$, and $2.5\%$, respectively, not the budget-indexed $\mathrm{ASR}@N$ of Table~\ref{tab:asr},
so Table~\ref{tab:semantic} should not be compared cell-by-cell with
Table~\ref{tab:asr}.

Table~\ref{tab:semantic} shows the outcome. Undefended, the cache is a
near-perfect oracle on every objective and backend (AUC $\ge0.99$; closed-world
ASR $96$--$99\%$ against a $10\%$ baseline). \NAME-Semantic returns all three
attacks to their random baselines (AUC $\approx0.5$, timing gap $\le5$\,ms) and
pads \emph{zero} benign traffic in our paraphrase and diverse-topic workloads,
because a scan must spread across many entities while benign traffic does not.
The defense principle therefore carries across cache layers.

\begin{table}[t]
\centering
\caption{Semantic-cache generalization. ASR baselines $50/10/2.5\%$; AUC $0.5$ = closed; $\Delta t$ = mean miss$-$hit gap (ms), $\approx\!0$ once equalized (small negatives are noise).}
\label{tab:semantic}
\resizebox{\columnwidth}{!}{
\begin{tabular}{ll cc c cc c}
\toprule
& & \multicolumn{3}{c}{\textbf{No Defense}} & \multicolumn{3}{c}{\textbf{\NAME-Semantic}} \\
\cmidrule(lr){3-5}\cmidrule(lr){6-8}
\textbf{Attack Objective} & \textbf{Backend} & ASR$\downarrow$ & AUC$\rightarrow$.5 & $\Delta t\downarrow$ & ASR$\downarrow$ & AUC$\rightarrow$.5 & $\Delta t\downarrow$ \\
\midrule
\multirow{3}{*}{Membership}
 & Qwen1.5-1.8B & 99.2 & 1.00 & 71.6  & 54.1 & 0.55 & -0.6 \\
 & Llama-3-8B   & 99.2 & 1.00 & 246.8 & 46.8 & 0.48 & -0.9 \\
 & DeepSeek-14B & 99.2 & 1.00 & 448.8 & 53.7 & 0.47 &  4.3 \\
\midrule
\multirow{3}{*}{Closed-world}
 & Qwen1.5-1.8B & 96.0 & 1.00 & 70.0  & 11.0 & 0.51 &  0.0 \\
 & Llama-3-8B   & 99.0 & 1.00 & 243.2 & 13.0 & 0.54 &  0.2 \\
 & DeepSeek-14B & 98.0 & 1.00 & 444.3 &  4.0 & 0.49 & -0.3 \\
\midrule
\multirow{3}{*}{Open-world}
 & Qwen1.5-1.8B & 62.0 & 0.99 & 68.8  &  3.0 & 0.48 & -0.2 \\
 & Llama-3-8B   & 58.0 & 0.99 & 240.3 &  2.0 & 0.50 & -0.1 \\
 & DeepSeek-14B & 54.0 & 0.99 & 437.6 &  1.0 & 0.47 & -0.3 \\
\bottomrule
\end{tabular}}
\end{table}

\subsection{Robustness to Adaptive Adversaries}
\label{sec:adaptive}
We assume a strong white-box adversary who knows \NAME's full design and parameters (the scoring rule, two-phase policy, equalization target, and decay constant $\gamma$) and adapts accordingly, but is not granted access to \NAME's internal state: it cannot see whether a response was a genuine miss or a padded hit, nor read a prefix's current score or phase. It attacks the residuals the mechanism leaves behind in two ways: passively mining distribution shapes under lockdown, and actively manipulating the temporal decay.

\noindent\textbf{Multivariate Distribution Distinguisher.}
Rather than compare mean latencies, the attacker takes $R\in\{10,20\}$ measurements per candidate, extracts an 8-feature vector (mean, median, min, q10, q25, q75, q90, std), and trains an L2-regularized logistic regression to separate hits from misses, evaluated over $200$ controlled hit/miss pairs with $5$-fold cross-validation. We report the effective AUC $\max(\text{AUC},1{-}\text{AUC})$. Against an undefended server the channel is wide open and sharpens with repetition exactly as Corollary~\ref{cor:repeat} predicts: CV AUC rises from $0.807$ ($R{=}10$) to $0.962$ ($R{=}20$, Table~\ref{tab:adaptive}), while against \NAME the corollary's growth term $1-(1-\deq)^R$ stays negligible at the measured $\widehat{\delta}_{\mathsf{eq}}$.

An $8$-feature logistic model ($R{=}20$ repeats: mean, median, and tail quantiles of TTFT) is trained to separate padded hits from natural misses. Against \NAME's clamp emitter its effective AUC is $0.53\pm0.01$ over six runs. We report $0.53$ rather than ``chance'' because the small gap is real: by the estimator of Section~\ref{sec:security} it is a measured equalization gap $\widehat{\delta}_{\mathsf{eq}}=2(0.53)-1\approx0.06$, a residual of at most $0.06$ bits per probe. This does not compose into a usable channel: by Theorem~\ref{thm:composition} recovering one high-entropy token position needs on the order of $1/0.06\approx17$ padded probes per bit, each still bounded by cache residency (Equation~\ref{eq:survive}). The honest summary is near chance with a small, measured residual.

\noindent\textbf{Whole-trace distinguisher.} The TTFT distinguisher above tested
only first-token features, so it left Assumption~\ref{as:regime}(A2) untested:
that the post-first-token schedule, i.e.\ the \emph{inter-token latencies}
(ITLs), leaks nothing about the cache bit beyond $o_1$. We therefore build a
stronger distinguisher that observes the \emph{entire} streaming trace, the
first-token latency \emph{and} every inter-token gap, by requesting $16$ tokens
and timestamping each. Crucially, to isolate a pure \emph{timing} channel from a
content artifact, the attacker always sends the \emph{same} probe $X$; we vary
only the cache state, priming $X$ for a hit and priming a sibling $P\Vert\text{wrong}$
(a realistic partial miss: the shared prefix $P$ stays warm, only the probed
candidate is cold) for a miss. The generated text is therefore identical and any
residual is pure timing. We extract TTFT features, ITL features (per-position
gaps, dispersion, and lag-1 autocorrelation), and the raw per-trace vector, and
train gradient-boosted and random-forest classifiers in addition to logistic
regression, reporting $5$-fold effective AUC. This is a strictly stronger family
than the $8$-feature TTFT logistic model, and it tests A2 head-on.

Table~\ref{tab:itl} evaluates this distinguisher on Llama-3-8B (excluding Qwen1.5-1.8B due to its inherently weak baseline channel; its undefended EarlyBird is only $6\%$). The results yield two critical insights. First, the ITL channel is highly exploitable: without defenses, the ITL-only and whole-trace classifiers achieve AUCs of $0.90$ and $0.99$, respectively. This proves that a naive defense that only pads the first token while exposing the original decode cadence remains deeply vulnerable. 
Second, \NAME effectively neutralizes this threat. Its whole-stream time-shift holds the ITL-only ($0.53$), whole-trace ($0.56$), and raw per-trace ($0.58$) classifiers near random chance. Because \NAME resamples padded hits from the current miss load regime, the preserved ITLs carry no incremental signal beyond the TTFT-only baseline ($0.58$), empirically validating Assumption A2. We observe identical success on DeepSeek-14B, the model with the widest undefended separation (whole-trace AUC $0.98$). \NAME collapses its AUCs down to $0.60$--$0.63$.

\begin{table}[t]
\centering
\caption{Content-controlled whole-trace distinguisher (effective AUC, $0.5$ = closed; fixed probe, pure timing). \NAME holds TTFT, ITL, and whole-trace features at chance (A2).}
\label{tab:itl}
\resizebox{\columnwidth}{!}{
\begin{tabular}{ll cccc}
\toprule
\textbf{Model} & \textbf{Defense} & \textbf{TTFT} & \textbf{ITL-only} & \textbf{Whole trace} & \textbf{Per-trace} \\
\midrule
\multirow{2}{*}{Llama-3-8B}
 & No Defense     & $0.98$ & $0.90$ & $0.99$ & $0.99$ \\
 & \textbf{\NAME} & $\mathbf{0.58}$ & $\mathbf{0.53}$ & $\mathbf{0.56}$ & $\mathbf{0.58}$ \\
\midrule
\multirow{2}{*}{DeepSeek-14B}
 & No Defense     & \ITLdsNODEFttft & \ITLdsNODEFitl & \ITLdsNODEFboth & \ITLdsNODEFpt \\
 & \textbf{\NAME} & \ITLdsMIRttft & \ITLdsMIRitl & \ITLdsMIRboth & \ITLdsMIRpt \\
\bottomrule
\end{tabular}}
\end{table}

\noindent\textbf{Low-and-slow evasion.} A patient adversary can instead wait out the odometer's time-decay and spread \emph{distinct} candidates so each is served in the unpadded grace phase. We analyze this spread-candidate strategy and the exact-repeat variant, and show that both collapse to the guessing baseline in deployment, in Appendix~\ref{app:sweeps}.

\subsection{Ablation Study}
\label{sec:ablation}

This section justifies our primary structural design choice: the shape of the emission target. Extensive parameter sweeps for all scalar hyperparameters (the grace budget $G$, repeat weight $\rho$, clamp quantile $q$, and margin $\eta$), alongside evaluations of low-and-slow adaptive evasion, confirm the robustness of our default configurations and are detailed in Appendix~\ref{app:sweeps}.

\noindent\textbf{Emission target.}
The clamp of Equation~\ref{eq:clamp} over-pads every hit to the miss tail. A tempting alternative is to \emph{match} the miss distribution so a fast hit is padded only to a typical miss, $\tau=F_{\mathsf{miss}}^{-1}(U)$. We implemented and tested this family carefully, at $n{=}3$ runs of the multivariate distinguisher each (Table~\ref{tab:emitter}). Matching the full law ($U\sim\mathrm{Uniform}(0,1)$) is the cheapest emitter ($10$\,ms mean pad) but leaks (effective AUC $0.67$): in the adversarial setting genuine misses are cold full-prefill executions and are slower than a typical recent miss, so a full-law draw emits some hits \emph{faster} than the cold misses and the distinguisher reads a ``hit is faster'' residual. Restricting the draw to the upper tail (a quantile floor $q_0{=}0.7$) or replacing the empirical window with a fitted parametric log-normal tail both \emph{reduce} the residual but do not robustly remove it: they average AUC $0.55$--$0.58$ with a large run-to-run standard deviation ($\pm0.06$--$0.10$) and leak on some runs, because reconstructing the cold-miss upper tail from a finite online window is inherently unreliable. Only the clamp, which sidesteps estimation by over-padding to the tail by construction, holds at chance ($0.53\pm0.01$) in \emph{every} run. \NAME therefore pays the clamp's extra overhead in exchange for robust indistinguishability; a low-overhead distributional emitter with the same guarantee remains open.

\begin{table}[t]
\centering
\caption{Emitter ablation (effective AUC, $n{=}3$ runs). Only the clamp closes the channel in every run; distributional emitters cut overhead but leave a residual.}
\label{tab:emitter}
\resizebox{0.9\columnwidth}{!}{
\begin{tabular}{l cc}
\toprule
\textbf{Emitter} & \textbf{Eff.\ AUC} $\rightarrow$.5 & \textbf{Mean pad (ms)} $\downarrow$ \\
\midrule
\textbf{Clamp (\NAME default)}        & \textbf{0.53$\,\pm$.01} & $35$ \\
Full CDF ($q_0{=}0$)                  & 0.67$\,\pm$.04 & $10$ \\
Tail-conditioned ($q_0{=}0.7$)        & 0.55$\,\pm$.06 & $12$ \\
Parametric log-normal ($q_0{=}0.7$)   & 0.58$\,\pm$.10 & $23$ \\
\bottomrule
\end{tabular}}
\end{table}

\section{Related Work}
\label{sec:related-work}

\noindent\textbf{LLM Timing Side-Channel Attacks.}
Recent work shows that AI serving optimizations can become timing oracles. EarlyBird~\cite{early-bird} studies timing leakage in LLM serving and demonstrates prompt recovery from cache-dependent latency differences. InputSnatch~\cite{input-snatch} turns prefill-time differences into practical black-box input recovery attacks across model and hardware configurations. PromptPeek~\cite{prompt-peek} shows that KV-cache sharing in multi-tenant serving can leak prompts by probing candidate continuations and observing cache-related behavior. Complementing these attacks, auditing work~\cite{audit-PSA} measures prompt caching behavior in commercial APIs and shows that externally visible timing can reveal when cached prefixes are reused. Other LLM side channels exploit different observable signals, including token lengths in encrypted streams~\cite{weiss2024promptremotekeyloggingattack} and timing variations introduced by efficient inference mechanisms such as speculative decoding~\cite{remote-attacks}. \NAME targets the KV-prefix-cache timing channel specifically, and assumes the attacker observes the full streaming trace rather than TTFT alone.

\noindent\textbf{LLM Timing Side-Channel Defenses.}
The most direct defense is to prevent cross-tenant cache reuse. Cache partitioning~\cite{cache-partitioning} assigns each user an isolated prefix-cache namespace, eliminating cross-user timing interactions at the cost of losing shared-prefix benefits. Production systems can implement a related policy using cache salts or extra hash keys~\cite{vllmCacheSalt} so that only requests with the same trust boundary share cached blocks. These mechanisms are simple and strong, but they penalize benign workloads with common public templates or shared system prompts, the very traffic that caching is built to reuse.

Selective-sharing systems try to recover some performance by distinguishing sensitive from shareable cache entries. SafeKV~\cite{safeKV}, for example, combines privacy detection, unified public/private metadata, and runtime safeguards to restrict reuse of sensitive KV blocks while preserving safe sharing. Such designs narrow the blast radius of isolation, but cannot escape it: every prefix marked as sensitive by the classifier is still denied cross-tenant reuse, so benign users of that prefix pay the full recomputation cost. This makes the content classifier a single point of failure for both security (a missed prefix leaks completely) and performance (an over-broad label isolates benign traffic), and it leaves the attacker's defining behavior, the burst of distinct candidate probes under one prefix, entirely unexamined. \NAME instead keeps sensitive prefixes in the shared cache and reacts to that probing behavior, obfuscating only the emitted timing once a prefix's distinctive fan-out signature shows up.

\section{Conclusion}

In this paper, we introduced \NAME, mitigating KV Cache side-channels by shifting timing delays directly at the API gateway rather than pausing computation or enforcing strict spatial barriers. Supported by a conditional information-theoretic analysis and extensive evaluation, with designs such as correct sensitivity gating, consistent prefix metering, and well-calibrated equalization, \NAME attains security comparable to absolute cache separation against the probing attacks we evaluate, up to a small, measured residual and under stated assumptions, while adding near-zero latency on public traffic under high multi-tenant loads and paying only a bounded, quantified cost on sensitive traffic that is under active probing.

\newpage
\bibliographystyle{IEEEtran}
\bibliography{references}
\newpage
\appendices

\section{Deferred Proofs}
\label{app:proofs}
This appendix collects the proofs of the equalization results of Section~\ref{sec:security}; the statements, assumptions, and resulting parameter rules remain in the main text.

\begin{proof}[Proof of Lemma~\ref{lem:reduction} (Trace reduction)]
Under (A2) the joint density of $\Tr$ given $B=b$ factorizes as
$p_b(o)\,K(g \mid o)$ with the \emph{same} kernel $K$. Writing
$D(o) := p_{\hit}(o) - p_{\miss}(o)$ and using $K \ge 0$,
\begin{equation}
  \TV\bigl(\Lo^{\Tr}_{\hit}, \Lo^{\Tr}_{\miss}\bigr)
  = \tfrac{1}{2}\iint \bigl|D(o)\bigr| K(g \mid o)\, dg\, do
  = \tfrac{1}{2}\int \bigl|D(o)\bigr|\, do ,
\end{equation}
since $\int K(g \mid o)\, dg = 1$; the last expression is
$\TV(\Lo^{o}_{\hit}, \Lo^{o}_{\miss})$. The identity is exact:
post-first-token emissions carry \emph{no} additional information about $B$.
Without (A2) only ``$\ge$'' holds, and the excess is exactly the residual that an
ITL-aware distinguisher can mine.
\end{proof}

\begin{proof}[Proof of Lemma~\ref{lem:advantage} (Bounded advantage and capacity)]
\emph{(a)} For any rule $\psi$ the advantage is at most
\begin{equation*}
  \sup_A\bigl|\Lo^{\Tr}_{\hit}(A) - \Lo^{\Tr}_{\miss}(A)\bigr|
  = \TV\bigl(\Lo^{\Tr}_{\hit},\Lo^{\Tr}_{\miss}\bigr),
\end{equation*}
attained by the likelihood-ratio test (Le~Cam); Lemma~\ref{lem:reduction} and
(A3) bound it by $\deq$.

\emph{(b)} Let $\delta = \TV(\Lo^{o}_{\hit},\Lo^{o}_{\miss}) \le \deq$; the
cases $\delta \in \{0,1\}$ are immediate, so take $\delta \in (0,1)$. With
densities w.r.t.\ a dominating measure $\mu$, set
\begin{equation*}
  R = \frac{\min(p_{\hit},p_{\miss})}{1-\delta},\qquad
  P'_{b} = \frac{(p_{b} - p_{\bar b})_+}{\delta},
\end{equation*}
which are densities because $\int \min(p_{\hit},p_{\miss})\,d\mu = 1-\delta$
and $\int (p_b - p_{\bar b})_+ d\mu = \delta$, have disjoint supports, and give $p_{b} = (1-\delta)R + \delta P'_{b}$ for
both $b$. So we may sample equivalently: draw $Z \in \{c,d\}$ with $\Pr[Z=d]=\delta$
independent of $B$; if $Z=c$ draw $o_1 \sim R$, else $o_1 \sim P'_B$; then
$g_{2:n} \sim K(\cdot \mid o_1)$ by (A2). Revealing $Z$ only helps, so $I(B;\Tr) \le I(B;\Tr,Z) = I(B;Z) + I(B;\Tr \mid Z)$.
Here $I(B;Z)=0$ since $Z \perp B$; given $Z=c$, $o_1 \sim R$ regardless of
$B$ and $g_{2:n}$ depends on $B$ only through $o_1$, so
$I(B;\Tr \mid Z=c)=0$; given $Z=d$ bound trivially by $H(B)=h_2(\pi)$.
Averaging, $I(B;\Tr) \le \delta\,h_2(\pi) \le \deq$, and $h_2 \le 1$ gives
$C_{\mathsf{eq}} \le \deq$.
\end{proof}

\begin{proof}[Proof of Corollary~\ref{cor:repeat} (Repeated measurement)]
Couple $\Lo^{\Tr}_{\hit}$ and $\Lo^{\Tr}_{\miss}$ maximally per coordinate:
they agree with probability $1-\deq$ each, hence in all $R$ with probability
$(1-\deq)^R$, and the TV distance between the product laws is at most the
coupling-failure probability. For the mutual information, run $R$ independent
genies $Z_i$: $I(B;\Tr^{(R)}) \le \Pr[\exists i: Z_i=d]\,h_2(\pi)$.
\end{proof}

\begin{proof}[Proof of Theorem~\ref{thm:composition} (Odometer--equalization composition)]
By the chain rule \eqref{eq:chainrule},
$I(S^*;O_{1:k} \mid P) \le \sum_{i=1}^{k} C_i$ with $C_i$ the $i$-th query's
capacity; the attacker fixes each continuation, so $O_i$ depends on $S^*$
only through that probe's bit $B_i$, and data processing gives
$I(S^*;O_i \mid P,O_{<i}) \le I(B_i;O_i \mid P,O_{<i}) \le C_i$. A Phase-1
query is an unpadded oracle, so $C_i \le C_1$ (the binary idealization gives $C_1{\le}1$; as discussed after Theorem~\ref{thm:fanout}, a continuous TTFT can raise it to $C_1{=}O(\log L)$, which replaces $N_1$ by $N_1C_1$ below and only strengthens the case for a small $N_1$); a Phase-2 query has
$C_i \le \deq$ by Lemma~\ref{lem:advantage}(b), which requires the bound
for \emph{every} prior, not merely $\pi=\tfrac12$. Hence
$I(S^*;O_{1:k} \mid P) \le N_1 + (k-N_1)\deq$, while Fano \eqref{eq:fano}
gives $I \ge H(S^*\!\mid\!P) - h_2(\varepsilon) - \varepsilon\log_2(M-1)$;
solving for $k$ yields \eqref{eq:composition}. Each query's phase is
measurable with respect to the attacker-controlled history, so $N_1$ may be
taken at its worst case; novel branches contributed by other tenants only advance
lockdown, never delay it.
\end{proof}

\section{Performance Bounds}
\label{sec:perf-bounds}

The preceding results bound what an attacker can \emph{learn}; we now bound what the defense \emph{costs}, so that the security dial $(q,\eta,G)$ comes with an explicit price tag. Three quantities matter: the latency of benign traffic, the worst-case delay of a padded request, and the serving capacity retained relative to isolation. Throughout, $c_{\mathsf{ctrl}}$ denotes the constant control-plane cost of one odometer decision, $\pi_2\in[0,1]$ the fraction of requests emitted in Phase~2, and $\tau_q(L)=(1+\eta)\bigl(a+bL+(Q_q(R))_+\bigr)$ the clamp target of Equation~\ref{eq:tau-affine} for a length-$L$ request under lockdown.

\begin{theorem}[Per-request latency bound]\label{thm:latency}
Under \NAME: \emph{(a)} every Phase-0/1 request is emitted at $e_1-t_0=t_{\mathsf{actual}}+c_{\mathsf{ctrl}}$, i.e., benign and grace traffic is never padded and pays only the constant control-plane term; \emph{(b)} every Phase-2 request satisfies
\begin{equation}
e_1-t_0\;\le\;\max\{t_{\mathsf{actual}},\,\tau_q(L)\}+\sigma+c_{\mathsf{ctrl}},
\label{eq:latency-bound}
\end{equation}
so its added delay obeys $\Delta_{\mathsf{pad}}\le(\tau_q(L)+\sigma-t_{\mathsf{actual}})_+$ while every inter-token gap is preserved exactly; \emph{(c)} averaging over the mix, \NAME's expected TTFT inflation over No~Defense is at most
$c_{\mathsf{ctrl}}+\pi_2\,\mathbb{E}\bigl[(\tau_q(L)+\sigma-t_{\mathsf{actual}})_+\bigr]$.
\end{theorem}

\begin{proof}
(a) holds by construction of the phase gate; (b) follows from Equations~\ref{eq:timeshift}--\ref{eq:tau-affine} and the jitter bound $\epsilon\le\sigma$; (c) sums the two cases weighted by $\pi_2$.
\end{proof}

\noindent\textbf{Isolation parity.}
Equation~\ref{eq:latency-bound} yields a direct comparison point. A padded request is emitted no later than an $\eta$-margined upper quantile of a \emph{genuine} cold miss, which is exactly the physical latency that isolation defenses impose on every cross-boundary request via recomputation. \NAME's worst per-request latency therefore exceeds isolation's typical one by at most the margin factor $(1+\eta)$, the residual tail term $Q_q(R)$, and the jitter $\sigma$; unlike isolation, this worst case is charged only to the $\pi_2$ slice of traffic under active probing rather than unconditionally, and Appendix~\ref{sec:collateral} shows the pessimal-accounting realization of this bound ($87$ vs.\ $88$\,ms).

\begin{theorem}[Capacity dominance over isolation]\label{thm:capacity}
Model each request as GPU demand $w=d+p_{\mathsf{suf}}$, where $d$ is decode work and $p_{\mathsf{suf}}$ the prefill of the uncached suffix, and let the engine be work-conserving with maximal sustainable throughput $\lambda_{\max}=\mu/\mathbb{E}[w]$ for GPU capacity $\mu$. Because \NAME executes padding as a gateway timer on already-computed chunks and metering on the host CPU, it adds zero GPU demand, hence
\begin{equation}
\lambda_{\max}^{\text{\NAME}}=\lambda_{\max}^{\mathsf{NoDef}},
\qquad
\frac{\lambda_{\max}^{\text{\NAME}}}{\lambda_{\max}^{\mathsf{iso}}}
=1+\frac{\bar h\,\mathbb{E}[p_P]}{\mathbb{E}[w]}\;\ge\;1,
\label{eq:capacity}
\end{equation}
where $p_P$ is the prefill cost of the shared prefix and $\bar h$ the fraction of requests whose cross-boundary reuse isolation is forced to drop.
\end{theorem}

\begin{proof}
\NAME leaves the engine's admission, scheduling, and kernels untouched, so per-request GPU demand equals No~Defense's and the first equality follows from work conservation. Isolation converts each forbidden reuse into a recomputation, raising demand from $w$ to $w+p_P$ on the $\bar h$ fraction, so $\mathbb{E}[w_{\mathsf{iso}}]=\mathbb{E}[w]+\bar h\,\mathbb{E}[p_P]$ and the ratio follows from $\lambda_{\max}=\mu/\mathbb{E}[w]$.
\end{proof}

Equation~\ref{eq:capacity} makes the throughput gap of Figure~\ref{fig:goodput} predictable rather than empirical: the observed $1.6\times$ headroom on Llama-3-8B corresponds to a workload where forbidden prefix reuse accounts for $\bar h\,\mathbb{E}[p_P]/\mathbb{E}[w]\approx0.6$ of the undefended demand, and the gap widens monotonically with prefix length and hit rate, which is exactly what prefix caching is for.

\begin{proposition}[The $(q,\eta)$ dial: leakage vs.\ padding]\label{prop:dial}
Suppose the online estimate satisfies $\sup_x|\widehat F_{\mathsf{miss}}(x)-F_{\mathsf{miss}}(x)|\le\varepsilon_{\mathsf{est}}$, and let $p_{\mathsf{slow}}=\Pr[t^{\hit}_{\mathsf{actual}}>\tau_q(L)]$ be the probability that a physical hit overshoots the clamp under load. Then the clamp emitter's equalization gap obeys
\begin{equation}
\deq\;\le\;(1-q)\;+\;\varepsilon_{\mathsf{est}}\;+\;p_{\mathsf{slow}},
\label{eq:deq-bound}
\end{equation}
while the expected pad on a fast hit is
$\mathbb{E}[\Delta_{\mathsf{pad}}\mid\hit]=\tau_q(L)+\sigma/2-\mathbb{E}[t^{\hit}_{\mathsf{actual}}]$,
which is nondecreasing in both $q$ and $\eta$.
\end{proposition}

\begin{proof}
Condition on a Phase-2 prefix. A fast hit ($t\le\tau_q$) is emitted at $\tau_q+\epsilon$; a miss is emitted at $\max(t,\tau_q)+\epsilon$, i.e., at $\tau_q+\epsilon$ with probability $F_{\mathsf{miss}}(\tau_q)$ and on its natural tail otherwise. The two emitted laws therefore coincide except on an event of probability at most $\bigl(1-F_{\mathsf{miss}}(\tau_q)\bigr)+p_{\mathsf{slow}}$; since $\tau_q$ dominates the estimated $q$-quantile, $F_{\mathsf{miss}}(\tau_q)\ge q-\varepsilon_{\mathsf{est}}$, and a maximal coupling converts the event bound into the total-variation bound~\eqref{eq:deq-bound}. The common jitter $\epsilon$ does not affect the distance. Monotonicity of the pad is immediate from the definition of $\tau_q$.
\end{proof}

Proposition~\ref{prop:dial} closes the loop between the two analyses. Lemma~\ref{lem:advantage} \emph{assumed} an equalization gap $\deq$ (A3); Equation~\ref{eq:deq-bound} now bounds that gap by deployment-controlled quantities, and substituting it into Theorem~\ref{thm:composition} prices security in milliseconds: driving the per-probe capacity down to $(1-q)+\varepsilon_{\mathsf{est}}+p_{\mathsf{slow}}$ costs exactly the tail pad of Theorem~\ref{thm:latency}(b), and both move monotonically with the same knob. The bound is also consistent with measurement: at $q{=}0.97$ it predicts $\deq\lesssim0.03+\varepsilon_{\mathsf{est}}+p_{\mathsf{slow}}$, in line with the measured $\widehat{\delta}_{\mathsf{eq}}\approx0.06$ of Section~\ref{sec:adaptive}, whose excess over $0.03$ empirically attributes the residual to finite-window estimation and load-induced slow hits rather than to any structural leak of the emitter. Finally, the control-plane state is itself bounded: each metered prefix stores a decayed scalar and a branch-hash set, $O(1)$ plus $O(|B_P|)$ hashes, and $B_P$ may be replaced by a fixed-size sketch whose one-sided error only \emph{over}-counts novelty, a fail-safe direction that can tighten memory without weakening Equation~\ref{eq:kmax}.

\section{Experimental Setup Details}
\label{app:setup}
\noindent\textbf{Hardware and serving configuration.} All experiments run on a multi-GPU server with NVIDIA A40 GPUs. The parallel runner assigns each SGLang victim server an exclusive GPU; detector, proposer, and proxy helper processes run on separate helper GPUs when available. The victim model is served with longest-prefix-match scheduling, a single in-flight request, and token-granularity paging. These settings intentionally sharpen prefix-cache timing and LPM order effects, making the configuration adversary-favorable for security testing. 

\noindent\textbf{Models and Datasets.} We evaluate three models spanning three families and three parameter scales (1.8B, 8B, 14B): \textit{DeepSeek-R1-Distill-Qwen-14B}~\cite{deepseekai2025deepseekr1incentivizingreasoningcapability}, \textit{Meta-Llama-3-8B-Instruct}~\cite{llama3modelcard}, and \textit{Qwen1.5-1.8B-Chat}~\cite{qwen}.

Security experiments use a 500-sample corpus of confidential consultation prompts assembled entirely from public sources: patient names from U.S. Census given-name and surname lists~\cite{uscensus_2020_names}, medical conditions and questions from MedQuAD~\cite{BenAbacha-BMC-2019} (\repo{lavita/MedQuAD}), and a shared system prompt from \repo{gabrielchua/system-prompt-leakage}~\cite{chua_system_prompt_leakage}. Each prompt embeds a sensitive name and condition beneath a shared sensitive prefix. The overhead workload uses only public instruction-following prompts from \repo{tatsu-lab/alpaca} \cite{taori2023stanford}. Thus the overhead benchmark measures benign public-prefix reuse rather than mixed sensitive traffic.

\noindent\textbf{Defense configuration.} \NAME uses threshold $\theta=1.0$, decay $\gamma=0.995$/s, sensitivity weight $\omega=1.0$ on sensitive prefixes, SGLang pagedattention block size $B=1$ token, dual-track target margin $0.20$ at the $0.97$ quantile, and $10$\,ms uniform emission noise; this is the $G=1$ grace budget, so the first novel branch beneath a sensitive prefix passes freely and a rapid second candidate triggers lockdown.
 The $B=1$ setting is adversary-favorable and avoids relying on coarse physical KV blocks to suppress attacks.

\section{Confidence Intervals}
\label{app:asr-ci}
Every ASR reported in Table~\ref{tab:asr} is \emph{directly measured} (Section~\ref{sec:metrics}); this appendix quantifies its sampling variability. EarlyBird and InputSnatch probe their finite candidate pools in full (EarlyBird's $467$ dataset first-tokens, InputSnatch's full $500{\times}500$ name/condition dictionary), so ASR is measured at every budget. OptiLeak probes, for every sensitive token position, the $\le\!50$-candidate rank window containing the ground truth at its true rank out to rank $10{,}000$; because the co-submitted batch stays $\le\!50$ at any depth, the order channel is exercised directly at each budget without a rank-tail estimator.

\noindent\textbf{Bootstrap protocol.}
We compute $95\%$ percentile bootstrap intervals with $5{,}000$ resamples. EarlyBird and InputSnatch resample the $500$ evaluation samples with replacement. OptiLeak resamples at the sample/prompt granularity and carries all token positions from a sampled prompt together; this cluster bootstrap avoids treating tokens from one prompt as independent. An all-zero empirical channel has interval $[0,0]$, which should be read as no observed channel in this experiment rather than a population-level impossibility proof; for measured-zero \NAME cells we report a rule-of-three upper bound instead.

\begin{table}[htbp]
\centering
\caption{Headline ASR with $95\%$ percentile bootstrap CIs (\%): EarlyBird and OptiLeak at ASR@10k, InputSnatch end-to-end. \NAME measured-zero cells show a rule-of-three upper bound.}
\label{tab:asr-ci}
\resizebox{\columnwidth}{!}{
\begin{tabular}{llcc}
\toprule
\textbf{Model} & \textbf{Attack} & \textbf{No Defense} & \textbf{\NAME (Ours)} \\
\midrule
\multirow{3}{*}{DeepSeek-14B}
 & EarlyBird   & 92.4 [90.0, 94.6] & 0.2 [0.0, 0.6] \\
 & InputSnatch & 59.4 [55.0, 63.8] & 0.4 [0.0, 1.0] \\
 & OptiLeak    & 68.8 [66.0, 71.6] & 0.0 [0.0, 0.1] \\
\midrule
\multirow{3}{*}{Llama-3-8B}
 & EarlyBird   & 33.4 [29.4, 37.6] & 0.0 [0.0, 0.6] \\
 & InputSnatch & 34.2 [30.0, 38.4] & 0.4 [0.0, 1.0] \\
 & OptiLeak    & 81.8 [79.1, 84.3] & 0.0 [0.0, 0.1] \\
\midrule
\multirow{3}{*}{Qwen1.5-1.8B}
 & EarlyBird   & 6.2 [4.2, 8.4] & 0.0 [0.0, 0.6] \\
 & InputSnatch & 0.0 [0.0, 0.0] & 0.0 [0.0, 0.6] \\
 & OptiLeak    & 67.6 [65.6, 69.6] & 0.0 [0.0, 0.1] \\
\bottomrule
\end{tabular}}
\end{table}

\section{Benign Sharing Cost}
\label{sec:collateral}

What happens when a sensitive prefix is never attacked but legitimately shared by many users? To answer this, we stress-test the most pessimistic configuration on Llama-3-8B: a single global budget that does not distinguish between individual users. Under this setting, a sustained stream of distinct benign queries from different users looks mathematically identical to an attacker's fan-out probing. Consequently, the mechanism errs on the side of caution and triggers lockdown, padding $93\%$ of these legitimate requests. While this high padding rate represents a theoretical drawback of a purely global budget, it is fundamentally a deliberate fail-safe behavior: when the system cannot separate identities, it prioritizes privacy. 

More importantly, the actual penalty for this "false positive" is remarkably cheap. Because \NAME applies the delay strictly at the gateway, the padded request is still served from the shared GPU cache. The worst-case penalty is merely a bounded $\approx\!60$\,ms shift ($87$\,ms padded versus the native $25$\,ms hit). In other words, \NAME's absolute worst-case benign penalty costs exactly what isolation defenses charge unconditionally on every request. Furthermore, this global budget is merely a diagnostic floor. As Table~\ref{tab:collateral} demonstrates, enabling trust-scoped metering completely resolves this flaw. By assigning each authenticated user their own budget, the mechanism cleanly separates normal usage from malicious fan-out, immediately dropping the benign padding rate to $0\%$ and restoring native latency, while remaining fully capable of locking down cross-account Sybil attacks.

\begin{table}[t]
\centering
\caption{Benign-sharing stress test (Llama-3-8B, $G{=}1$, pessimal global budget): a pad is a small shift on a still-cached hit; scoped metering removes it. $^\dagger$Simulation.}
\label{tab:collateral}
\resizebox{0.8\columnwidth}{!}{
\begin{tabular}{l l c c}
\toprule
\textbf{Metering policy} & \textbf{Benign pattern} & \textbf{Padded} & \textbf{TTFT (ms)} \\
\midrule
Global budget & distinct continuations & $93\%$ & $87$ \\
Global budget & exact repeats ($\rho{=}0$) & $\mathbf{0\%}$ & $\mathbf{25}$ \\
\midrule
Trust-scoped$^\dagger$ & distinct continuations & $\mathbf{0\%}$ & $\mathbf{25}$ \\
\bottomrule
\end{tabular}}
\end{table}

\section{Sensitive-Shared Workloads}
\label{sec:sensshared}

This dynamic padding behavior directly translates to massive throughput advantages over selective isolation systems like SafeKV. We demonstrate this by comparing 64 benign tenants sharing a single sensitive prefix under varying loads. On purely public traffic, all systems perform identically ($87$--$89$\,ms). On sensitive traffic at a low load of $8$ req/s, the systems also appear matched: Cache Partition and SafeKV sit at $88$\,ms, while \NAME records $91$\,ms. This tight alignment highlights the intelligence of \NAME's dynamic estimator. Although \NAME actively pads these requests under the global budget stress test, the estimator targets the \emph{actual} concurrent miss latency. Because cache misses are relatively fast at low loads, \NAME's padded hit latency perfectly matches the physical recomputation latency of isolation defenses, so it effectively masks the padding penalty it applies.

However, the mechanisms diverge drastically as the load approaches the server's capacity at $16$ req/s. Strict isolation forces the GPU to redundantly recompute the sensitive prefix for every single tenant. This redundant prefill quickly saturates the GPU compute capacity, inflating TTFT to nearly $1.8$\,s (a severe $3\times$ degradation). In stark contrast, \NAME continues to reuse the shared cache at full GPU speed, merely applying its gateway-level padding constraint. This allows \NAME to maintain a highly responsive $611$\,ms TTFT, closely tracking the undefended $542$\,ms baseline. Ultimately, while \NAME introduces a minor gateway delay on sensitive hits to preserve security, this deliberate trade-off completely neutralizes the catastrophic GPU queuing collapse inherent to isolation defenses when throughput is contended.

\begin{table}[htbp]
\centering
\caption{TTFT (ms), $64$ tenants sharing one sensitive prefix (Llama-3-8B, $\text{MRR}{=}16$): near capacity, recomputation inflates isolation $\sim3\times$; \NAME tracks No~Defense.}
\label{tab:sensshared}
\resizebox{0.7\columnwidth}{!}{
\begin{tabular}{l c cc}
\toprule
\multirow{2}{*}{\textbf{Defense}} & \textbf{Public} & \multicolumn{2}{c}{\textbf{Sensitive-shared}} \\
\cmidrule(lr){2-2}\cmidrule(lr){3-4}
 & @8 req/s & @8 req/s & @16 req/s \\
\midrule
No Defense            & $87$ & $86$  & $542$ \\
\textbf{\NAME}        & $89$ & $91$  & $611$ \\
Cache Partition       & $88$ & $88$  & $1858$ \\
SafeKV (selective)    & $89$ & $88$  & $1777$ \\
\bottomrule
\end{tabular}}
\end{table}

\section{Security Under Real Batched Load}
\label{app:mrr}
Our main security numbers use a single in-flight request ($\text{MRR}{=}1$), the adversary-favorable setting that sharpens the timing signal, whereas a production server batches. We therefore re-run both channels against \NAME at $\text{MRR}{=}16$ with concurrent benign background load. The order channel stays at the guessing floor ($0\%$ channel success over $120$ targets): probes that land in the same batch are released through the same phase-gated barrier and fresh random permutation, and an attacker who forces serialization by occupying the run slots with slow filler merely recovers the $\text{MRR}{=}1$ regime that the batch-order equalization already closes. The magnitude/ITL channel likewise stays at chance: the whole-trace distinguisher of Table~\ref{tab:itl} stays within $0.53$--$0.58$ effective AUC against \NAME under $\text{MRR}{=}16$. This holds because the clamp targets an \emph{upper} quantile of the \emph{current} miss window, which only rises as batching thickens the miss tail, so added load can push the emitter toward over-padding but never below the concurrent miss law. Both the $\text{MRR}{=}1$ and the $\text{MRR}{=}16$ batched regimes are thus covered by our evaluation.

\section{Parameter Sweeps and the Low-and-Slow Adversary}
\label{app:sweeps}
This appendix collects the scalar parameter studies that fix \NAME's defaults and the analysis of the low-and-slow adaptive adversary that the grace budget admits (Section~\ref{sec:adaptive}). Each sweep varies one knob with the rest held at the main configuration.

\noindent\textbf{Grace budget $G$.} $G$ trades benign latency against the size of the unpadded window, and we justify the default $G{=}1$ by measuring \emph{both} axes on one model (Qwen1.5-1.8B; Table~\ref{tab:grace}). \emph{Latency:} $G{=}0$ pads every benign cache hit up to the miss target ($476$\,ms), whereas any $G\ge1$ serves it at the true hit latency ($\approx44$\,ms), a $>\!10\times$ reduction that \emph{saturates at $G{=}1$}, since $G{=}2$ and $G{=}3$ measure the same $\approx44$\,ms. \emph{Security:} the realistic spread-candidate top-1 ASR (Section~\ref{sec:adaptive}, no per-candidate oracle reset, $8$ candidates so a $12.5\%$ random baseline) stays at that baseline for every $G$ (averaged over \GRn\ seeds), so the wider window carries no realizable leakage. The trade-off is thus one-sided past $G{=}1$: the full latency benefit is already captured at $G{=}1$, while the unpadded window, and hence the theoretical exposure, grows linearly with $G$. The principled choice is the smallest budget that saturates the benefit, $G{=}1$: it avoids $G{=}0$'s $>\!10\times$ benign-latency penalty and takes on none of the enlarged window of $G\ge2$. We retain $G{=}0$ as an option for providers who want to pad even first accesses (the only setting that also closes the stacked-oracle bound of Table~\ref{tab:adaptive}).

\begin{table}[t]
\centering
\caption{Grace-budget trade-off (Qwen1.5-1.8B, \GRn{} seeds): benign hit latency saturates at $G{=}1$; spread-candidate ASR stays at the $12.5\%$ baseline for every $G$.}
\label{tab:grace}
\resizebox{0.92\columnwidth}{!}{
\begin{tabular}{l cccc}
\toprule
\textbf{Grace budget} & $G{=}0$ & $\mathbf{G{=}1}$ & $G{=}2$ & $G{=}3$ \\
 & \footnotesize(pad-all) & \footnotesize(default) & & \\
\midrule
Benign hit latency (ms) $\downarrow$ & $476$ & $\mathbf{44}$ & $42$ & $46$ \\
Spread-candidate ASR (\%) & \GRa & \GRb & \GRc & \GRd \\
\bottomrule
\end{tabular}}
\end{table}

\noindent\textbf{Repeat weight $\rho$.} We ablate $\rho$ with the exact-repeat evasion: an attacker seeds a candidate, waits for the score to decay, and re-issues the \emph{identical} prompt hoping for an unpadded hit ($200$ trials, $R{=}5$; Table~\ref{tab:rho}). Realistically this already fails (effective AUC $0.50$) through \emph{self-cache pollution}, the seed caches the attacker's own wrong candidate and destroys the hit/miss contrast. Even under an idealized oracle that removes that pollution and keeps the victim's secret resident, which does reopen the channel ($0.975$), setting $\rho{=}1$ to charge exact repeats forces the harvest probe back into lockdown and closes it ($0.52$). We default to $\rho{=}0$ (free retries for benign duplicate traffic) because the oracle is not realizable under cache eviction (Section~\ref{sec:adaptive}), and expose $\rho{=}1$ for providers unwilling to rely on residency.

\begin{table}[t]
\centering
\caption{Repeat-weight ablation (exact-repeat evasion, effective AUC). Self-pollution already neutralizes it; the idealized oracle is closed by charging repeats ($\rho{=}1$).}
\label{tab:rho}
\resizebox{0.7\columnwidth}{!}{
\begin{tabular}{l c c}
\toprule
\textbf{Setting} & $\boldsymbol{\rho}$ & \textbf{Eff.\ AUC} $\rightarrow$.5 \\
\midrule
Realistic (self-pollution)      & $0$ & $\mathbf{0.501}$ \\
Idealized oracle                & $0$ & $0.975$ \\
Idealized oracle, charged       & $1$ & $\mathbf{0.516}$ \\
\bottomrule
\end{tabular}}
\end{table}

\noindent\textbf{Emitter quantile $q$ and margin $\eta$.} The clamp target is set
by the miss quantile $q$ and the safety margin $\eta$ (Section~\ref{sec:estimator});
to check that the default $q{=}0.97,\eta{=}0.20$ is not a delicate operating point,
we sweep both and measure the residual channel (the $8$-feature distinguisher's
effective AUC) and the mean pad (Table~\ref{tab:paramsweep}). Across $q\in\{0.90,0.97\}$ and
$\eta\in\{0.10,0.20,0.30\}$ the residual stays near chance ($0.51$--$0.58$) while
the pad rises gently and monotonically with the margin ($24\to33$\,ms). Security
is thus not a knife-edge in these knobs: within this range the channel stays
closed and only the overhead moves, so $q{=}0.97,\eta{=}0.20$ is a reasonable
operating point rather than a tuned one. The trend is the expected one: more
padding buys marginally more robustness at a small, predictable latency cost, and
a provider can slide along it without reopening the channel.

\begin{table}[t]
\centering
\caption{Clamp quantile $q$ and margin $\eta$ sweep (Qwen1.5-1.8B): the residual stays near chance while the pad moves modestly, with no knife-edge.}
\label{tab:paramsweep}
\resizebox{0.82\columnwidth}{!}{
\begin{tabular}{l cc}
\toprule
\textbf{Config} & \textbf{Eff.\ AUC} $\rightarrow$.5 & \textbf{Mean pad (ms)} $\downarrow$ \\
\midrule
$q{=}0.97,\ \eta{=}0.20$ (default) & $0.58$ & $28$ \\
$q{=}0.90,\ \eta{=}0.20$           & $0.52$ & $27$ \\
$q{=}0.97,\ \eta{=}0.10$           & $0.54$ & $24$ \\
$q{=}0.97,\ \eta{=}0.30$           & $0.51$ & $33$ \\
\bottomrule
\end{tabular}}
\end{table}

\noindent\textbf{Low-and-slow time-decay evasion.} The second strategy waits for the odometer score to decay before probing. Re-issuing the \emph{exact} same candidate is self-defeating and, even under an oracle, is closed by the repeat weight $\rho$; we treat this as the $\rho$ ablation (Section~\ref{sec:ablation}). The more dangerous variant spreads \emph{distinct} candidates over time, which we analyze here.

\noindent\emph{Spread-Candidate:} the attacker avoids self-pollution by spacing \emph{distinct} candidates over time, waiting for the score to decay below $\theta$ between probes so each is served in Phase~1. Our $G{=}1$ default admits one such unpadded probe per decay window. We report its worst case in Table~\ref{tab:adaptive}, and we stress that the $76.5\%$ figure ($\rho{=}0$; $25.5\%$ at $\rho{=}1$) is an artifact of three oracle assumptions stacked together, not a channel that exists in deployment. Concretely, it requires simultaneously that (a) the odometer score is reset to exactly zero before \emph{every} candidate, which forces the attacker to idle out a full multi-minute decay per probe (Equation~\ref{eq:decay-time}); (b) throughout those cumulative hours the victim's \emph{exact} secret prefix stays resident in the shared cache; and (c) the attacker's own probes never pollute that state. Assumptions (a) and (b) are directly contradictory under real load: the same idle time that decays the score is far more than enough for competing multi-tenant traffic to LRU-evict the victim's prefix, at which point every delayed probe is a natural miss and recovers nothing. This is exactly the spatio-temporal dilemma of Equation~\ref{eq:survive}, and a provider makes it explicit by choosing $\gamma$ so the per-probe wait exceeds the cache residency time $\tau_{\max}$. Even setting residency aside, the knobs \NAME already exposes each close the window independently: charging exact repeats ($\rho{=}1$) drops the oracle ASR to $25.5\%$, and $G{=}0$ pads the first probe and eliminates the unpadded window by construction. We therefore treat the spread-candidate rows as a labeled upper bound that quantifies the cost of the grace budget, not as a realizable attack.

\begin{table}[htbp]
\centering
\caption{White-box adaptive attacks (Qwen1.5-1.8B). The distinguisher stays at chance against the clamp; spread-candidate rows are a stacked-oracle bound, closed by $\rho{=}1$ or $G{=}0$.}
\label{tab:adaptive}
\resizebox{0.9\columnwidth}{!}{
\begin{tabular}{l c c c}
\toprule
\textbf{Configuration} & $\mathbf{R}$ & \textbf{Eff.\ AUC} $\rightarrow$.5 & \textbf{Top-1 ASR} $\downarrow$ \\
\midrule
\multicolumn{4}{l}{\textit{Multivariate Distinguisher}} \\
No Defense                       & 10 & 0.807 & -- \\
No Defense                       & 20 & 0.962 & -- \\
\textbf{\NAME\ (clamp)}          & 20 & \textbf{0.53$\,\pm$.01} & -- \\
\midrule
\multicolumn{4}{l}{\textit{Spread-Candidate Scan ($R{=}5$, baseline $\approx 0.51\%$)}} \\
Oracle ($\theta=1.0, \rho=0$) & 5 & -- & 76.5\% \\
Oracle ($\theta=1.0, \rho=1$) & 5 & -- & 25.5\% \\
\bottomrule
\end{tabular}}
\end{table}

\section{Limitations}
\label{sec:limitations}

\NAME makes deliberate trade-offs, and we state its boundaries plainly.

\noindent\textbf{Sensitivity Classification.} Lemma~\ref{lem:advantage} relies on correctly identifying sensitive prefixes. While our evaluation uses a lexical detector for benchmark reproducibility, \NAME's architecture does not inherently depend on text scanning. In real-world deployments, providers can adopt lightweight embedding-based sensitivity token detectors, or a fail-closed policy using existing application-level metadata (e.g., enterprise tenant IDs, private document flags, or proprietary endpoints) to route requests to the protected path ($\omega_{\mathsf{sens}}>0$) by default. Under this paradigm, content scanning serves merely as an optional performance whitelist to safely bypass the odometer for known-public templates, completely eliminating the risk of missing obfuscated, paraphrased, or multilingual secrets.

\noindent\textbf{Benign sharing of a metered prefix.} Under the pessimal single-global-budget accounting, a sustained stream of distinct benign continuations beneath one fail-closed prefix is mostly padded (Appendix~\ref{sec:collateral}). The pad is a bounded gateway shift on a request still served from the shared cache (no recomputation, no throughput loss), so this worst case matches what isolation charges every request unconditionally; exact repeats are never charged, trust-scoped accounting removes the padding for authenticated traffic entirely, and $G$ is a tunable dial.

\noindent\textbf{Estimator calibration.} The miss target $\tau$ is tracked online; during cold-start warmup \NAME falls back to conservative default miss timing rather than leaving Phase~2 unprotected. The threat model allows dummy traffic and cache flushes (Section~\ref{sec:threat-model}), which shift the miss distribution the estimator tracks, but the estimator's structure leaves such manipulation little to work with. Shifting the distribution \emph{up} only raises $\tau$: more padding, never less. Shifting it \emph{down} is structurally impeded three times over. The miss track admits only zero-cached-token observations, whose latency is bounded below by a genuine full prefill (partially cached ``warm'' requests are routed to the hit track and cannot dilute the miss window), and the affine fit normalizes away length games such as short-prompt floods. The target couples a margined affine center with the upper quantile of its recent residuals, so depressing it requires dominating essentially the entire recent pure-miss window, and the slow EMA tracks ($\beta{=}0.9995$) that seed and backstop the fit resist rapid displacement. And the genuine concurrent misses against which a padded hit would be compared are drawn from the same load regime the attacker's own flooding creates, so the reference moves with the window; the non-stationary $\text{MRR}{=}16$ runs of Appendix~\ref{app:mrr} exercise exactly this shifting-mix regime without reopening the channel. Hardening the window further with robust statistics is straightforward if a deployment wants defense in depth.

\noindent\textbf{Distributed deployment.} Our prototype keeps the odometer in a single tokenizer manager. A multi-replica deployment shares this state so that probes spread across replicas draw on one budget; because the state is small and additive (a decayed score and a branch-hash set per prefix), it maps directly onto the shared low-latency stores or per-prefix sticky routing that multi-replica gateways already employ for rate limiting. Quantifying the cross-replica read cost on the critical path is routine engineering we have not yet performed.

\noindent\textbf{Hardware and scale.} Our evaluation runs on NVIDIA A40 GPUs, single-GPU serving, and models up to $14$B. We do not evaluate A100/H100 hardware, tensor- or pipeline-parallel multi-GPU serving, or $70$B/MoE models, where prefill/decode timing, batch composition, and cache pressure differ. Because \NAME lives entirely in the gateway/control plane and adds no GPU work, we expect the mechanism to carry over, but the specific overhead and $\widehat{\delta}_{\mathsf{eq}}$ numbers should be read as characteristic of this regime, not of every stack out there.

\noindent\textbf{Semantic-cache generalization.} Section~\ref{sec:semantic} shows the \emph{principle} transfers to embedding-similarity caches; it is a generalization of the KV-cache mechanism, not a full port of the KV-cache attacks. Our evaluation there uses a $40$-topic medical corpus with real sentence embeddings and simulated hit/miss timing; real (rather than transplanted) semantic-cache attacks against a production embedding cache are still open for future work.

\noindent\textbf{Threat scope.} \NAME is purpose-built to neutralize API-visible timing side-channels arising from prefix-cache reuse, which currently represent the most accessible and scalable vector for cross-tenant prompt recovery. Consequently, it is not designed to be a panacea for all prompt recovery vulnerabilities. Specifically, \NAME does not address non-timing side-channels, such as analyzing encrypted network traffic bursts, monitoring token-length emission sequences or direct KV cache inversion attacks that extract tokens from raw KV cache tensors.

\section{Future Work}
\label{sec:future}
Several directions follow directly from the limitations above.

\noindent\textbf{Robust sensitivity gating.} Replacing the lexical detector with a learned or entropy-based gate that scores a prefix by the conditional entropy and PII-likelihood of its continuations, rather than by membership in a word list, would extend metering to paraphrased, obfuscated, and multilingual secrets and remove the benchmark-overfitting concern.

\noindent\textbf{Entropy-adaptive grace budget.} Equation~\ref{eq:union} ties the safe grace budget to the secret's min-entropy, which suggests granting more grace (and thus lower overhead) to provably hard-to-guess prefixes by estimating $H_{\infty}(S^*\!\mid\!P)$ online. The obstacle is that the serving model's next-token distribution measures linguistic likelihood, not sensitivity, and cannot lower-bound a stronger fine-tuned adversary's prior, so a naive estimate can grant unsafe grace. A safe version would use a conservative lower-confidence estimate and cap the result at the global budget $G$, $G_P=\min(G,\lfloor\alpha\,2^{\hat H_{\infty}^{\mathrm{LCB}}}\rfloor)$, so estimation error can only tighten protection; we leave a careful instantiation to future work.

\noindent\textbf{Extending threat coverage.} Addressing non-timing leakage channels, such as token counts, network traffic, or direct KV cache inversion, would broaden the protection scope of \NAME. Techniques like differential privacy, traffic shaping, or secure multi-party computation could be explored to mitigate these additional non-timing threats in future systems.
\end{document}